\documentclass[a4paper,12pt]{article}
\pdfoutput=1
\usepackage[affil-it]{authblk}
\usepackage[utf8]{inputenc}
\usepackage[T1]{fontenc}
\usepackage{lmodern}
\usepackage{graphicx,subfigure,multicol}
\usepackage{mathenv,amsmath,mathtools,amsfonts,amssymb}
\usepackage{makeidx,comment,multirow}
\usepackage{array,multirow,makecell}
\usepackage{geometry}
\usepackage{grffile}
\usepackage{cite}
\title{Statistical mechanics of the 3D axi-symmetric Euler equations in a Taylor-Couette geometry}

\author{Simon Thalabard$^1$ \thanks{\texttt{simon.thalabard@cea.fr}}
, Bérengère Dubrulle$^1$ and Freddy Bouchet$^2$}
\affil{$^1$ Laboratoire SPHYNX, Service de Physique de l'Etat Condensé,
DSM, CEA Saclay, CNRS URA 2464, 91191 Gif-sur-Yvette, France\\
$^2$
Laboratoire de Physique de l’Ecole Normale Supérieure de Lyon, Université de Lyon and CNRS, 46, Allée d'Italie, F-69007 Lyon, France
}
\begin{document}

\maketitle
\newtheorem{result}{Result}[section]
\newcommand{\dsp}{\displaystyle}

\newcommand{\mA}{\mathcal A}
\newcommand{\mC}{\mathcal C}
\newcommand{\mD}{\mathcal D}
\newcommand{\mE}{\mathcal E}
\newcommand{\mF}{\mathcal F}
\newcommand{\mG}{\mathcal G}
\newcommand{\mH}{\mathcal H}
\newcommand{\mN}{\mathcal N}
\newcommand{\mO}{\mathcal O}
\newcommand{\mP}{\mathcal P}
\newcommand{\mQ}{\mathcal Q}
\newcommand{\mS}{\mathcal S}
\newcommand{\mX}{\mathcal X}
\newcommand{\mW}{\mathcal W}

\renewcommand{\ln}{\log}

\newcommand{\klD}{\mathfrak D}
\newcommand{\slD}{\mS}
\newcommand{\mSK}{{{\mathfrak S}_K}}
\newcommand{\Ak}{\{A_k\}}
\newcommand{\Xk}{\{X_k\}}
\newcommand{\nhC}{\mC_{\text{n.h.}}}
\newcommand{\tf}{\tilde{f}}
\newcommand{\tF}{\tilde{F}}
\newcommand{\tmF}{\tilde{\mathcal{F}}}
\newcommand{\tmD}{\tilde{\mathcal{D}}}
\newcommand{\ty}{\tilde{y}}

\newcommand{\KL}{Kullback-Leibler }
\newcommand{\pprime}{{\prime\prime}}

\newcommand{\condhel}{partial circulations }

\newcommand{\vD}{\left| \mathcal D \right|}
\newcommand{\dx}{\textbf{dx}\,}


\renewcommand{\c}{(c)}
\renewcommand{\o}{(o)}
\newcommand{\m}{(m)}
\renewcommand{\d}{\text{d}}
\newcommand{\x}{\left({\bf x}\right)}
\newcommand{\xprime}{\left({\bf x^\prime}\right)}
\newcommand{\xnod}{\left({\bf x_0}\right)}
\newcommand{\dxnod}{\left|{\bf d x_0}\right|}
\newcommand{\xij}{\left({\bf x_{N,ij}}\right)}
\newcommand{\xijp}{\left({\bf x_{N,i^\prime j^\prime}}\right)}

\newcommand{\dsx}{\,d\sigma_N\,\otimes d\xi_N \,}
\newcommand{\dpmn}{\,\d\mP_{M,N} \,}
\newcommand{\dpmntor}{\,\d\mP_{N}^{tor,E} \,}
\newcommand{\dpmnpol}{\,\d\mP_{M,N}^{pol,E} \,}
\newcommand{\dpmnc}{\,\d\mP_{M,N}( \mC) \,}

\newcommand{\probx}{p_M\left(\xi,{\bf x } \right)}
\newcommand{\probkx}{p_{k,M}\left(\xi,{\bf x } \right)}
\newcommand{\pstar}{p^\star_M \left(\xi,{\bf x} \right)}
\newcommand{\pstark}{p^\star_{M,k} \left(\xi,{\bf x} \right)}
\newcommand{\pstarE}{p^{\star,E}_M \left(\xi,{\bf x} \right)}

\newcommand{\pfun}{Z^{\star}_M \xij}
\newcommand{\pfunx}{Z^{\star}_M \x}
\newcommand{\iM}{_{M}}
\newcommand{\lM}{^{(M)}}
\newcommand{\iL}{_{(l)}}

\newcommand{\gibbs}{{gibbs}}
\newcommand{\micro}{{micro}}
\newcommand{\pol}{{pol}}
\newcommand{\tor}{{tor}}
\newcommand{\tot}{{tot}}
\newcommand{\coef}{\dfrac{\left|\mathcal{D}\right|}{N^2}}

\newcommand{\pg}{\paragraph{}}
\newcommand{\spg}{\subparagraph{}}

\newcommand{\isp}[1]{[\![#1]\!]}

\newcommand{\simon}[1]{{#1}}

\begin{abstract}
In the present paper, microcanonical measures for the dynamics of three dimensional (3D) axially symmetric turbulent flows with swirl in a Taylor-Couette geometry are defined,  using an analogy with a long-range lattice model.  We compute the relevant physical quantities and argue that two kinds of equilibrium regimes exist, depending on the value of the total kinetic energy. For low energies, the equilibrium flow consists of a purely swirling flow whose toroidal profile depends on the radial coordinate only. For high energies, the typical toroidal field is uniform, while the typical poloidal field is organized into either a single vertical jet or a large scale dipole, and exhibits infinite fluctuations. This unusual phase diagram comes from the poloidal fluctuations not being bounded for the axi-symmetric Euler dynamics, even though the latter conserve infinitely many ``Casimir invariants''.  This shows that 3D axially symmetric flows can be considered as intermediate between 2D and 3D flows. 
  
\end{abstract}

\setcounter{tocdepth}{2}
\tableofcontents
\section{Introduction}

\pg
Statistical mechanics provides powerful tools to study complex dynamical systems in all fields of physics.  However, it usually proves difficult to apply classical statistical mechanics ideas to turbulence problems. The main reason is that many statistical mechanics theories relie on equilibrium or close to equilibrium results, based on the microcanonical measures. Yet, one of the main phenomena of classical three dimensional (3D) turbulence is the anomalous dissipation, namely the existence of  an energy flux towards small scales that remains finite in the inertial limit of an infinite Reynolds number. This makes the classical 3D turbulence problem an intrinsic non-equilibrium problem. Hence, microcanonical measures have long been thought to be irrelevant for turbulence problems. 

\pg
A purely equilibrium statistical mechanics approach to 3D turbulence is actually pathological. Indeed, it leads for any finite dimensional approximation to an equipartition spectrum, which has no well defined asymptotic behavior in the limit of an infinite number of degrees of freedom \cite{bouchet2011statistical}. This phenomena is related to the Rayleigh-Jeans paradox of the equilibrium statistical mechanics of classical fields \cite{pomeau1994statistical}, and is a sign that an equilibrium approach is bound to fail. This is consistent with the observed phenomena of anomalous dissipation for the 3D Navier-Stokes and suspected equivalent anomalous dissipation phenomena for the 3D Euler equations.  

The case of the 2D Euler equations and related Quasi-Geostrophic dynamics is a remarkable exception to the rule that equilibrium statistical mechanics fails for classical field theories. In this case, the existence of a new class of invariants -- the so-called ``Casimirs'') and  among them the enstrophy -- leads to a completely different picture. Onsager first anticipated this difference when he  studied the statistical mechanics of the point vortex model, which is a class of special solutions to the 2D Euler equations \cite{onsager1949statistical,eyink2006onsager}. After the initial works of Robert, Sommeria and Miller in the nineties \cite{miller1990statistical,robert1991statistical,robert1992relaxation} and subsequent work \cite{michel1994statistical,jordan1997ideal,ellis2004statistical,majda2006nonlinear,bouchet2010invariant}, it is now clear that microcanonical measures taking into account all invariants exist for the 2D Euler equations. These microcanonical measures can be built through finite dimensional approximations. The finite dimensional approximate measure has then a well defined limit, which verifies some large deviations properties -- see for instance \cite{potters2012sampling} for a recent simple discussion of this construction. The physics described by this statistical mechanics approach is a self-organization of the flow into a large scale coherent structure corresponding to the most probable macrostate.

\pg
The three dimensional axi-symmetric Euler equations describe the motion of a perfect three dimensional flow, assumed to be symmetric with respect to rotations around a fixed axis. Such flows have additional Casimir invariants, which can be classified as ``toroidal Casimirs'' and ``helical Casimirs'' (defined below). By contrast with the 2D Euler equations,  the Casimir constraints do not prevent the vorticity field to exhibit infinitely large fluctuations, and it is not clear whether they can prevent an energy towards smaller and smaller scales, although it has been stated that the dynamics of such flows should lead to predictable large scale structures \cite{monchaux2006properties}. Based on these remarks, the three dimensional axi-symmetric Euler equations seem to be an intermediate case between 2D and 3D Euler equations, as previously suggested in \cite{leprovost2006dynamics,naso2010statistical}. It is then extremely natural to address the issue of the existence or not of non-trivial microcanonical measures. 

\pg
\simon{The present paper is an attempt to write down a full and proper statistical mechanics equilibrium theory for axially symmetric flows in  the microcanonical ensemble, directly  from first principles, and releasing the simplifying assumptions previously considered in the literature. Examples of such assumptions included  either a non-swirling hypothesis \cite{mohseni2001statistical,lim2003coherent}, an hypothesis that the equilibria are governed by restricted sets of ``robust invariants'' \cite{leprovost2006dynamics} or a deterministic treatment of the poloidal field \cite{leprovost2006dynamics,naso2010statistical,naso2010statistical2}.
 Those simplifying hypothesis have proved extremely fruitful in giving a phenomenological entropic description of ring vortices or of the  large-scale coherent structures observed in swirling flows generated in von K\'{a}rm\'{a}n setups \cite{monchaux2006properties,monchaux2007mecanique}. As far as the 3D axi-symmetric Euler equations ares concerned though, those treatments were in a sense not completely satisfying. Besides, whether they should lead to relevant invariant measures is not clear. } 

\pg
To derive the  axi-symmetric equilibrium measures,  we define approximate microcanonical measures on spaces of finite dimensional approximations of axially symmetric flows, compatible with a formal Liouville theorem. As the constrained invariant subspace of the phase space is not bounded, we also have to consider an artificial cutoff $M$ on the accessible vorticity values. From these approximate microcanonical measures, we compute the  probability distribution of poloidal and toroidal part of the velocity field.  The microcanonical  measure of the 3D axi-symmetric equations  is defined as a weak limit of  sequences of those finite dimensional approximate microcanonical measures, when the cutoff $M$ goes to infinity. More heuristically stated, we will show that finite dimensional approximations of the Euler equations can be mapped onto a long-range lattice model whose thermodynamic limit, obtained in the limit of the lattice  mesh going to zero, defines a microcanonical measure of the Euler equations. We prove that the limit exists and that it describes non-trivial flow structures.

\pg
\simon{
Our treatment of the poloidal fluctuations yields a very thought-provoking phase diagram, which describes the existence of two different regimes of equilibrium. The control parameter is the total kinetic energy. When the kinetic energy is low, the equilibrium flow is characterized by a positive (microcanonical) temperature. In this regime, the typical field is  essentially toroidal and is stratified as it depends on the radial coordinate only. When the kinetic energy is higher than a threshold value,  the toroidal field is uniform and the poloidal field is both non-vanishing and non-trivial. While the typical poloidal field is dominated by large scales, the equilibrium state  exhibits infinitely large fluctuations and is  non-gibbsian. As a result, the  microcanonical temperature is infinite. In both regimes, it is \emph{found} that the average field is a steady state of the axi-symmetric Euler equations, formally stable with respect to any axially symmetric perturbation}. \\



\subparagraph{}
\simon{
The paper is organized as follows. 
In Section 2, we introduce the axi-symmetric Euler equations together with their associated  Casimir functions. We then relate the axi-symmetric equilibrium measures to microcanonical ensemble described in the thermodynamic limit of a well-defined long-range lattice model model. Although the main result of our paper concerns the case where all the Casimirs are taken into account, we find it enlightening and pedagogic to consider before hands some toy equilibria obtained by deliberately ignoring  all the correlations between the toroidal and the poloidal fields induced by the presence of the helical casimirs. The analysis is carried out in Section 3.  Those correlations are restored in Section 4.  We find out that the phase diagram obtained in the simplified case of section 3 is exactly the one that describes the full problem. We discuss about the physical content of our results in Section 5.}

\section{Mapping the axi-symmetric Euler equations onto a spin model}

In this section, we introduce the axi-symmetric Euler equations and their invariants. We discretize them in physical space, and observe that the corresponding equilibrium statistical model is described by a  lattice model in which the ``spins'' can be pictured as point-wise Beltrami vortices (to be defined below)  with non local interactions. We argue that there exists a natural microcanonical thermodynamic limit for the spin model. It describes a continuous axially symmetric field,  and induces an invariant measure of the axi-symmetric Euler equations. 

\subsection{Axi-symmetric Euler equations and dynamical invariants}

\subsubsection{Equations}
\pg
The starting point of the study are the Euler equations for incompressible flows inside a domain $\mD$  in between  two concentric cylinders of height $2h$, with internal radius $R_{in}$ and outer one $R_{out}$, and whose volume we write  $ \vD  = 2h\pi \left( R_{out} ^2 - R_{in}^2 \right) $.
The Euler equations read : 
\begin{equation}
 \partial_t \bf{v} + {\bf v}. {\bf \nabla}  {\bf v} = -{\bf \nabla}  {\bf p} \text{~~and~~} {\bf \nabla}.  {\bf v}=0.
 \label{eq:3Deuler}
 \end{equation}

We use cylindrical coordinates $(r,\theta,z)$ and consider axi-symmetric flows within a cylindrical geometry. Those flows are defined through their three velocity components $v_r$, $v_{\theta}$ and $v_z$ depending on $r$ and $z$ only. Instead of the usual velocity variables $\bf{v}$, it proves convenient to write the Euler equations for axi-symmetric flows in terms of a toroidal field $\dsp \sigma = rv_\theta$, together with a poloidal field $\dsp \xi = \dfrac{\omega_\theta}{r} = \dfrac{\partial_z v_r - \partial_r v_z}{r}$.
It  also proves convenient to use the coordinate $y= \dfrac{r^2}{2}$ instead of $r$, and we write $\dx = \d y \d \theta \d z$ the infinitesimal cylindrical volume element at position $\x=(y,\theta,z)$.  

\spg
In the present study, we focus on velocity fields which are $2h$-periodic along the vertical direction and which satisfy an impermeability boundary condition on the two cylindric walls, namely $\left. \bf{v}.\bf{n} \right|_{\partial \mD}=0$  -- with $\bf n$ the unit  vector normal to the boundary ${\partial \mD}$.  Since the flow is incompressible  (${\bf \nabla} . {\bf v} = 0$), we know (Helmholtz decomposition) that there exists a periodic stream function $\psi$ and a constant $C$ such that $(2y)^{\frac{1}{2}}v_r = -\partial_z \psi+C$ and $v_z = \partial_y\psi$. The impermeability boundary condition imposes that $C=0$. Besides, without lack of generality, $\psi$ can be chosen  such that it is vanishing on both the inner and the outer walls.
\footnote{$\psi$ is defined up to a constant. Since $\psi$ takes a constant value on both the outer and on the inner walls, one of those constants can be set to $0$ without lack of generality. Then, using Equation (\ref{eq:3Deuler}) and the boundary conditions, one observes that the quantity  $\dsp {\mathcal M}_z = (2h)^{-1}\int_\mD \d y \d z v_z =   \left. \psi \right|_{R^2_{in}/2} - \left. \psi \right|_{R^2_{out}/2}$ is a conserved by the Eulerian dynamics be it or not axi-symmetric. Therefore, we can choose to consider the referential in which ${\mathcal M}_z$ is zero, and in which  $\left. \psi \right|_{R^2_{in}/2} = \left. \psi \right|_{R^2_{out}/2}=0$.}
The fields $\xi$ and $\psi$ are then related through
\begin{equation}
-\xi = \Delta_\star \psi = \dfrac{1}{2y}\partial_{zz} \psi + \partial_{yy}\psi,  \text{~~and~~} \psi =0 \; \text{ on both the inner and the outer walls.} 
\label{eq:xi-psi}
\end{equation}

Therefore, prescribing  both the toroidal and the poloidal field $(\sigma,\xi) $ also completely prescribes the three dimensional axially symmetric velocity field $(v_r,v_\theta,v_z)$ -- and vice-versa. 

\spg
The axi-symmetric Euler equations for the $(\sigma ,\xi)$ variables read \cite{szeri1988nonlinear,leprovost2006dynamics}
\begin{align}
& \partial_t \sigma + \left[\psi,\sigma\right] =0 \text{~~and~~} \partial_t \xi + \left[\psi,\xi\right] = \partial_z \dfrac{\sigma^2}{4y}.
\label{eq:axi}
\end{align}

The inner-brackets represent the advection terms and are defined by $\left[f,g\right]=\partial_yf\partial_zg - \partial_zf\partial_yg$.
\simon{We note that the toroidal field is not only transported by the poloidal field but also exerts a feedback on the poloidal evolution equation. It behaves as an active scalar. The feature is not an artifact of the cylindrical geometry. The generation of poloidal vorticity  by the toroidal field can be interpreted as the effect of the centrifugal forces acting on the fluids, which is akin but not completely equivalent to the effect of the Lorentz force on the kinetic vorticity field in 2D magneto-hydrodynamics \cite{Vladimirov1997}, or buoyancy effects in the 2D Boussinesq equations \cite{abarbanel1986nonlinear}}.
\spg
Unless stated otherwise, we will assume from now on that $R_{in}$ is non-zero ($R_{in}>0$), hereby considering a so-called ``Taylor-Couette'' geometry. 

\subsubsection{Dynamical invariants}
\pg
It is straightforward to check that the kinetic energy  $\dsp E= \dfrac{1}{2}\int_\mD \dx {\bf v}^2 $  is a conserved quantity of the axi-symmetric Euler equations ~(\ref{eq:axi}). The kinetic energy  can be written in terms of the fields $\sigma$ and $\xi$ as 
\begin{equation}
 E = \dfrac{1}{2} \int_\mD \dx \left[ \dfrac{\sigma^2}{2y} + \xi \psi\right].
\label{eq:axienergy}
\end{equation}

\spg
As a consequence of Noether theorem (for the relabelling symmetry) and the degeneracy of its Hamiltonian structure (\cite{morrison1998hamiltonian,szeri1988nonlinear}), the axi-symmetric  Euler equations have infinitely many Casimir invariants. They fall into two families: the Toroidal Casimirs $C_f$ and the Helical Casimirs $H_g$, defined  by 
\begin{equation}
  C_f = \int_\mD \dx f\left( \sigma \right) \; \text{  and  } \;  H_g =  \int_\mD \dx \xi g\left( \sigma \right), \label{eq:Casimirs} 
\end{equation}
where $f$ and $g$ can be any sufficiently regular functions. 

\spg
Note that the well-known invariants of the incompressible Euler equations correspond to specific choices for the functions $f$ and $g$. The conservation of the usual helicity $H=\int_\mD \dx {\bf v}.{\bf \omega}$ is for instance recovered by setting   $g(x)\equiv 2x$ in equation (\ref{eq:Casimirs}). Setting $f(x)\equiv x $ gives the conservation of the  $z$-component of the angular momentum. Setting $g(x) \equiv 1$ gives the conservation of the circulation of the velocity field along a closed loop following the boundary of a meridional plane.

\subsection{Dynamical invariants seen as geometrical constraints}
\pg
We can give an alternative, more geometric, description of the Casimirs constraints (\ref{eq:Casimirs}). We introduce the indicator function $\mathbf{1}_{B(x)}$. This function takes value $1$ if $B(x)$ is true and $0$ otherwise. Now, given a value $q$ for the toroidal field, let us set $f\equiv g \equiv \mathbf{1}_{\sigma\x \le q}$ in equation (\ref{eq:Casimirs}). Doing so, we obtain the specific ``Toroidal Casimirs''  $\mC_q(\sigma)=\int_\mD\dx \mathbf{1}_{\sigma\x \le q}$ together with the specific ``Helical Casimirs'' $\mH_q(\sigma,\xi)=\int_\mD\dx \xi\x \mathbf{1}_{\sigma\x \le q}$.  

\spg
$\mC_q$ represents the area of $\mD$ where the toroidal field is lower than a prescribed value $q$. $\mH_q$ can be interpreted as the poloidal circulation on the contour of the domain corresponding to $\mC_q$. Deriving  $\mC_q$ and $\mH_q$ with respect to $q$, we find that the distribution of the poloidal field $\mA_q =\dfrac{1}{\vD} \dfrac{\d \mC_q}{\d q}$ together with the \condhel  $\mX_q =\dfrac{1}{\vD} \dfrac{\partial \mH_q}{\partial q}$ are dynamical invariants of the axi-symmetric equations.

\spg
The conservations of the all the areas $\mA_q$ together with that of all the \condhel $\mX_q$ is in fact equivalent to the conservations of the whole set of Casimirs -- Toroidal and Helical --  since for sufficiently regular functions $f$ and $g$ we can write $C_f$ and $H_g$ as 
\begin{equation}
 C_f\left[\sigma\right] = \vD \int_\mathbb{R}\d q \mA_q\left[\sigma\right] f(q)  \text{ and }  H_g\left[\sigma,\xi\right] = \vD \int_\mathbb{R}\d q \mX_q\left[\sigma,\xi\right] g(q).
\end{equation}

\spg
Now, consider a discrete toroidal distribution, say $\dsp f(\sigma) = \sum_{k=1}^K \dfrac{A_k}{\vD}  \mathbf{1}_{\sigma = \sigma_k}$. Let $\mathfrak{S}_K =\{\sigma_1,\sigma_2...\sigma_K\}$ be the discretized set of possible values for the toroidal field. In this simplified yet general situation, the conservation of the Casimirs is equivalent to the conservation of the $K$ areas and $K$ \condhel :

\begin{equation}
 \mA_k\left[\sigma\right] =  \int \dx \mathbf{1}_{\sigma\x=\sigma_k} \;  \text{ and } \; \mX_k \left[\sigma,\xi\right] =  \int \dx \xi \mathbf{1}_{\sigma\x=\sigma_k} \label{eq:alterCasimirs}.
\end{equation}

\simon{
Let us emphasize here that considering  the toroidal field as a discrete set of ``toroidal patches'' is totally consistent with the ideal  axi-symmetric equations. It is completely analogous to the vortex patch treatment of the vorticity field in 2D, on which more  details can be found for example in \cite{robert1991statistical,miller1990statistical}.}
\subsection{Analogy with an  ``axi-symmetric'' long-range lattice model}

\subsubsection{Discretization of the fluid}
\begin{figure}[htb]
 \includegraphics[width=0.24\textwidth,trim=0.6cm 0.5cm 1cm 0cm, clip]{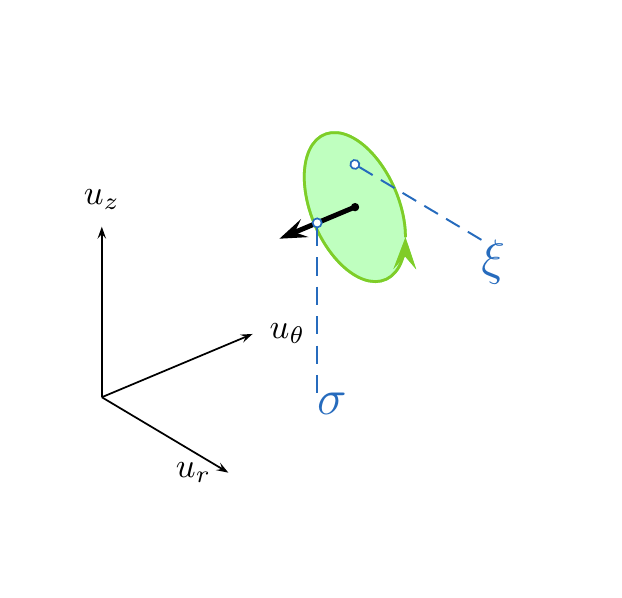}
 \includegraphics[width=0.24\textwidth,trim=2cm 0.5cm 1cm 0cm, clip]{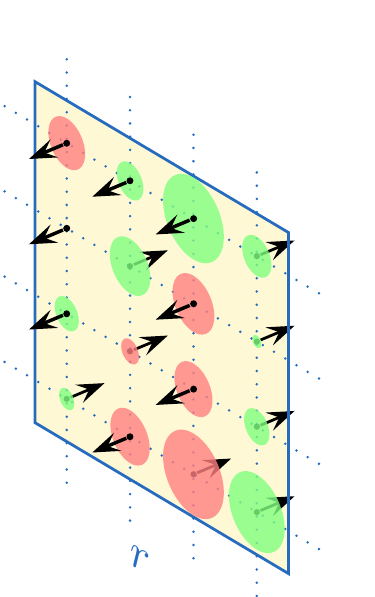}
 \includegraphics[width=0.24\textwidth,trim=8cm 1cm 8cm 0cm, clip]{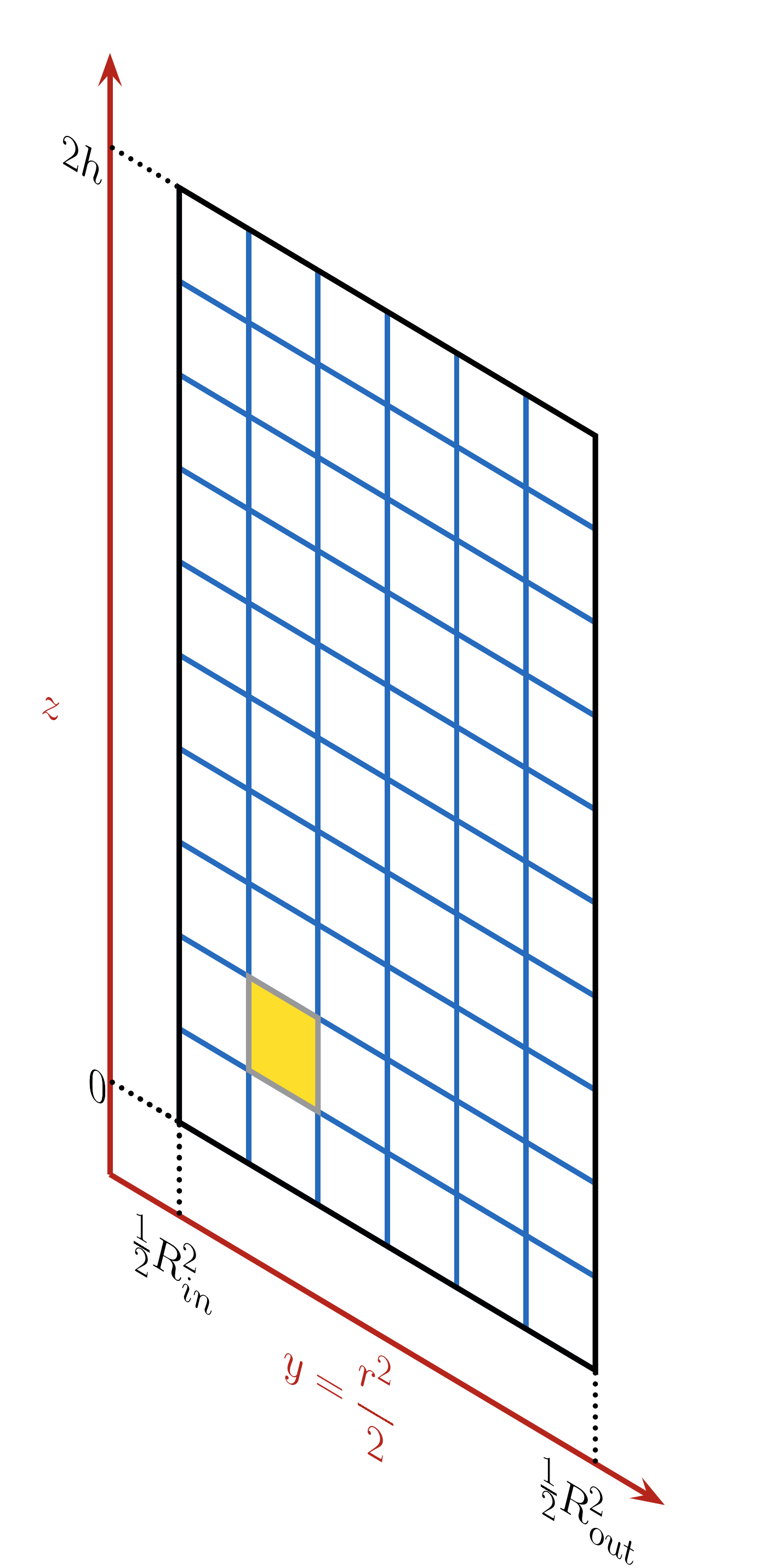}
 \includegraphics[width=0.24\textwidth,trim=26cm 10cm 26cm 5cm, clip]{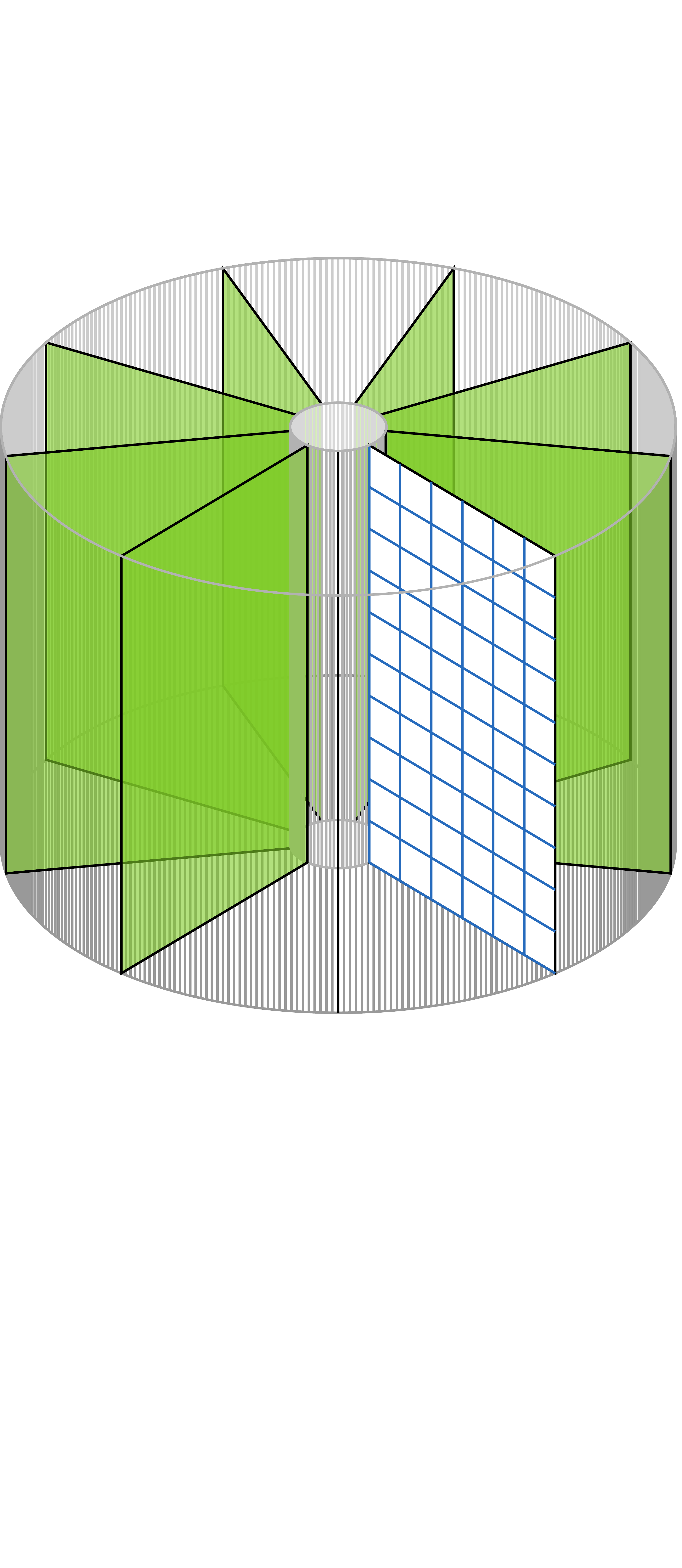}
\caption{Discretization of the axi-symmetric Euler equations onto an assembly of Beltrami spins (Impressionistic view). For each Beltrami spin, we represent the toroidal degree of freedom by an arrow, and the poloidal degree of freedom by a circle whose radius is proportionnal to the amplitude of the poloidal field.  Red (green) circles denote negative (positive) vorticy. }
\label{fig:sketch_disc}
\end{figure}

Let us  cut a slice of fluid along a meridional plane $ \mP $, and draw a $N\times N$ regular lattice on it. We can consider a discretization of the toroidal field and the poloidal field $\dsp (\sigma_N,\xi_N) = \left( \sigma_{N,ij}, \xi_{N,ij} \right)_{1\le i,j\le N} $. Each node of the grid corresponds to a position $\xij$ in the physical space, on which there exist a two-degree-of-freedom object that we refer to as an elementary ``Beltrami spin''. One degree of freedom is related to the toroidal field,  while the other is related to the poloidal field. The discretization procedure is sketched on Figure \ref{fig:sketch_disc}. It it simply the axi-symmetric extension to the construction developed in the 2D case in 
\cite{miller1990statistical,ellis2000large}.

\spg
We associate to every spin configuration a discretized version of the axi-symmetric energy (\ref{eq:axienergy}), that is discretized into the  sum of a toroidal energy and a poloidal energy, namely 

\begin{align}
& \mathcal{E}[\sigma_{N},\xi_{N}]  = \mathcal{E}_{tor}[\sigma_{N}] +\mathcal{E}_{pol}[\xi_{N}] \\
& \text{with } \mathcal{E}_{tor}[\sigma_{N}]=\dfrac{1}{4}\coef\hspace{-0.3cm}\sum_{(i,j)\in \isp{1;N}^2} \hspace{-0.3cm}\dfrac{\sigma_{N,ij}^2}{y_{i}}
 \; \text{  and  } \;\mathcal{E}_{pol}[\xi_{N}]=\dfrac{1}{2}\dfrac{\vD}{N^4} \hspace{-0.3cm}\sum_{\substack{(i,j)\in \isp{1;N}^2\\(i^\prime,j^\prime)\in \isp{1;N}^2}} \hspace{-0.3cm} \xi_{N,ij}G_{iji^\prime j^\prime}\xi_{N,i^\prime j^\prime}. 
\label{eq:discreteenergy}
\end{align}
$G_{iji^\prime j^\prime}$ denotes a discretized version of the Green operator $-\left(\Delta_\star\right)^{-1}$ with vanishing boundary conditions on the walls and periodic conditions along the vertical direction.

\spg
We now introduce the discretized counterparts of the  Casimir constraints (\ref{eq:alterCasimirs}) as
\begin{align}
\mA_k \left[\sigma_{N}\right] = \coef\hspace{-0.3cm}\sum_{(i,j)\in \isp{1;N}^2}\hspace{-0.3cm} \mathbf{1}_{\sigma_{N,ij}=\sigma_k} \; \text{ and } \; \mathcal{X}_{k}[\sigma_{N},\xi_{N}] =\coef\hspace{-0.3cm}\sum_{(i,j)\in \isp{1;N}^2}\hspace{-0.3cm} \xi_{N,ij} \mathbf{1}_{\sigma_{N,ij}=\sigma_k} \label{eq:discreteconstraints}.
\end{align}

Here, the indicator function $\mathbf{1}_{\sigma_{N,ij}=\sigma_k}$ is the function defined over the $N^2$ nodes of the grid, that takes value 1 when  $\sigma_{N,ij}=\sigma_k$ and 0 otherwise.  Let us also write the discrete analogue of the total poloidal circulation as $\dsp \mathcal{X}[\sigma_{N},\xi_{N}]=\sum_{k=1}^K \mathcal{X}_{k}[\sigma_{N},\xi_{N}]$. 

\begin{figure}[htb]
 \includegraphics[width=0.24\textwidth,trim=1cm 0.5cm 1cm 0cm, clip]{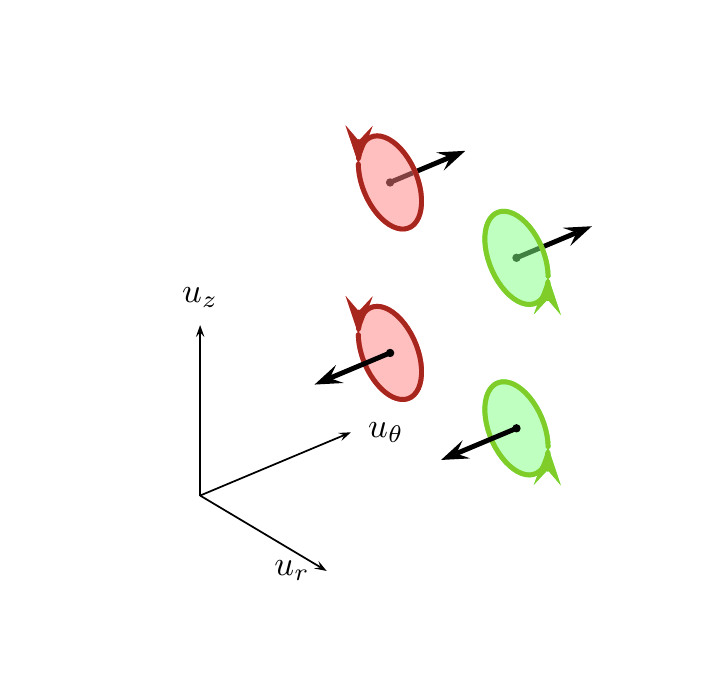}
 \includegraphics[width=0.24\textwidth,trim=1cm 0.5cm 1cm 0cm, clip]{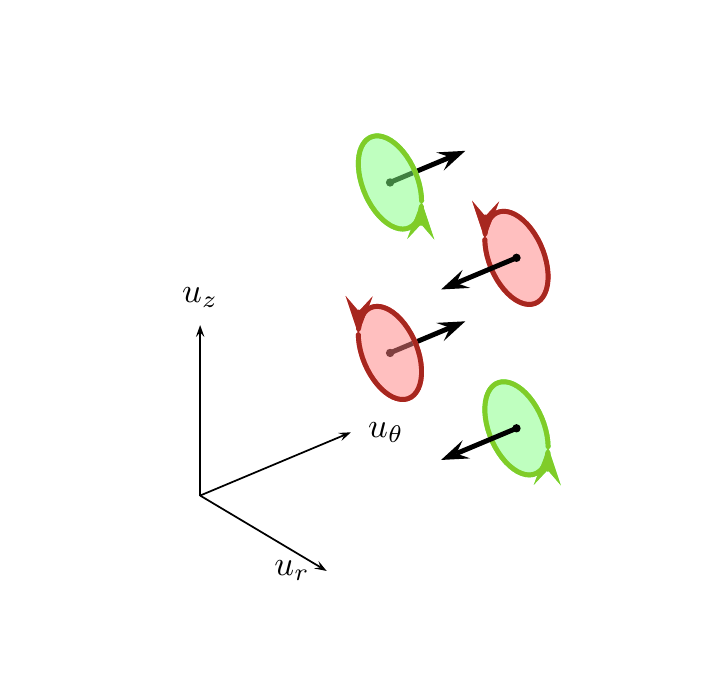}
 \includegraphics[width=0.24\textwidth,trim=1cm 0.5cm 1cm 0cm, clip]{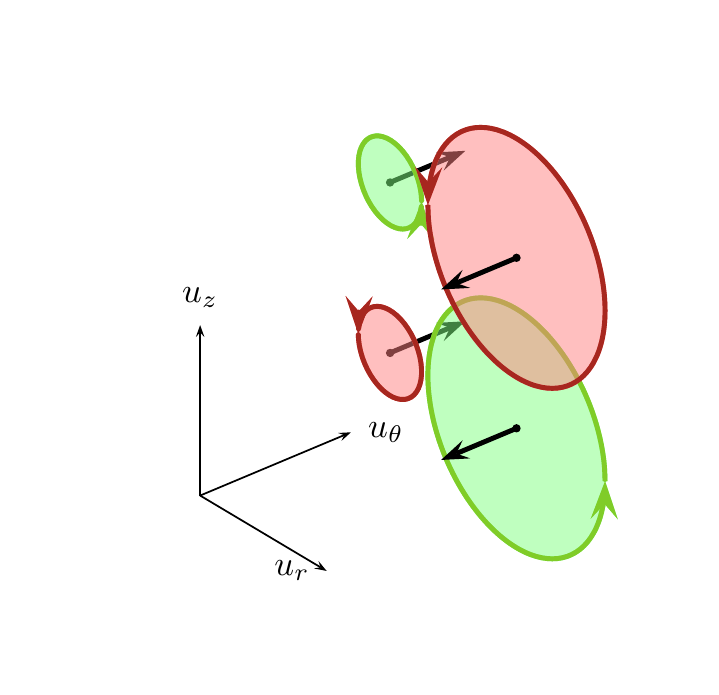}
 \includegraphics[width=0.24\textwidth,trim=1cm 0.5cm 1cm 0cm, clip]{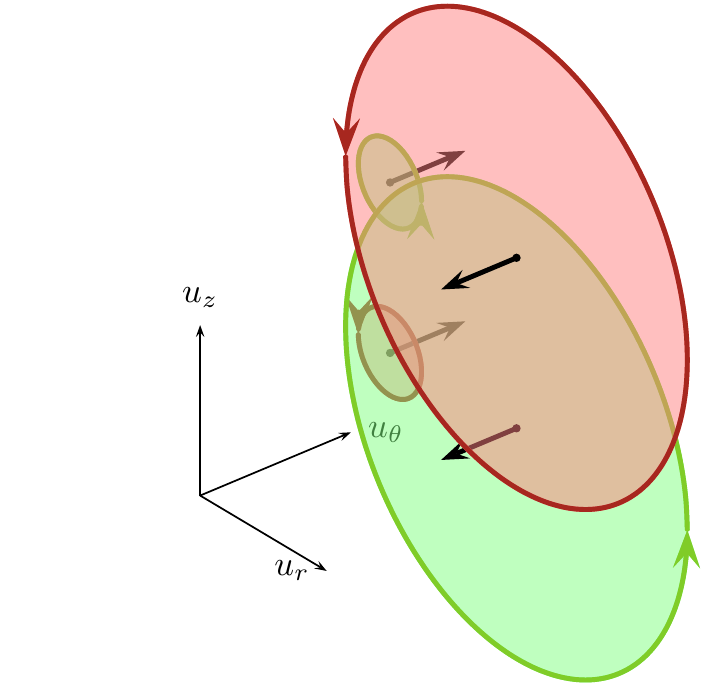}
 \caption{An assembly of four Beltrami Spins satisfying the same constraints on their Toroidal Areas and Poloidal Partial Circulations.}
\label{fig:sketch_toy}
\end{figure}

\spg
To make the constraints more picturesque, we have sketched on Figure \ref{fig:sketch_toy} different configurations of  an assembly of four Beltrami spins with two toroidal patches ($K=2$) and symmetric toroidal  levels ($\mathfrak{S}_2=\{-1, 1\} $). Each toroidal area occupies half of the domain : $A_1=A_{-1}=\dfrac{\vD}{2}$. The poloidal circulations conditioned on each one of the patches  are also zero : $X_1=X_{-1}=0$.

\subsubsection{The (helical) axi-symmetric microcanonical measure}
\simon{
The basic idea behind the construction of the microcanonical measure  is to translate the dynamical constraints imposed by the axi-symmetric ideal dynamics onto a well defined ``microcanonical ensemble''.} To do so, we consider the set $\mC$ of  $2K+1$ constraints given by 
\begin{equation}
\mC = \{E, \Ak_{1\le k\le K}, \Xk_{1\le k\le K}\}.
\end{equation}
Given $N$, we define the configuration space $\mG_N(E,\{A_k\},\{X_k\})\subset \left(K \times {\mathbb R} \right)^ {N^2}$  as the space of all the spin-configurations $\left(\sigma_N,\xi_N\right)$  that are such that $\dsp E \le \mE\left[\sigma_N,\xi_N\right] \le E+\Delta E $ and $\dsp \forall 1\le k \le K,\, \mA_k\left[\sigma_N\right] = A_k$, and $\mX_k\left[\sigma_N,\xi_N\right] = X_k$. As will be clear later on, the number of configurations increases exponentially with $N$. Then, in the limit of large $N$, due to this large deviation behavior, the microcanonical measure will not depend on $\Delta E$.\\
  
\simon{The salient properties of the present axi-symmetric lattice model stem from the lack of a natural bound for the poloidal degrees of freedom. Were we to define uniform measures directly on each one of the configuration spaces $\mG_N$, we would end up with trivial measures, as each one of the $N^2$ poloidal degrees of freedom can span the entire ${\mathbb R}$ range. }
To deal with this issue, we therefore  introduce  bounded ensembles $\mG_{M,N} $ made of the spin-configurations of $\mG_N$ that satisfy $({\sup_{ij}}\left|\xi_{N,ij} \right| \le M )$.    
For every ensemble $\mG_{M,N}$, we can then define a $M,N$ dependent  microcanonical measure $\d\mP_{M,N}$ together with a $M,N$ dependent microcanonical average $<>_{M,N}$ by assigning a uniform weight to the spin configurations in $\mG_{M,N}$. The construction of  $\d\mP_{M,N}$ and $<>_{M,N}$ is explicitly carried out in sections (\ref{ssub:microptilde}) and (\ref{ssub:micropfull}).  
\spg

 $M$ plays the role of an artificial \emph{poloidal cutoff}. \emph{A priori}, it  has no physical meaning  and is not prescribed by the axi-symmetric dynamics. It is natural to let it go to infinity.
  The present paper aims at building a thermodynamic limit by letting successively  $(N\to\infty)$ and $(M\to \infty)$ for this set of microcanonical measures,  and to describe this limit.  We will refer to this measure as the (helical) axi-symmetric measure.\\
  
Let us emphasize, that the two limits $(N\to\infty)$ and $(M\to \infty)$ most probably do not commute. We argue that the relevant limit is the limit  ($N\to\infty$) first. Taking the limit  $(N\to\infty)$ first, we make sure that we describe a microcanonical measure that corresponds to the dynamics of a continuous field (a fluid). The microcanonical measure at fixed $M$ then corresponds to an approximate invariant measure, for which the maximum value of the vorticity is limited. Such a fixed $M$ measure could be relevant as a large, but finite time approximation if the typical time to produce large values of the vorticity is much longer than the typical time for the turbulent mixing. Finally, for infinite time, we recover the microcanonical measure by taking the limit $(M\to\infty)$.  For these reasons, we think that the physical limit is the limit $(N\to\infty)$ first.\\

As for the physics we want to understand out of it, it is the following.  Consider  an assembly of Beltrami spins with a given energy $E$. What is the fraction of $E$ that typically leaks into the toroidal part and into the poloidal part ?  What does a typical distribution of Beltrami spins then look like ?  

\subsection{How is the axi-symmetric microcanonical measure related to the axi-symmetric Euler equations ?}
\subparagraph{}
Interpreting the invariants as geometrical constraints on a well- defined assembly of spin-like objects has allowed us to map the microcanonical measure of discretized hydrodynamical fields and invariants onto an long-range, ``Beltrami Spin'', lattice model. Taking the thermodynamical limit ($N \rightarrow  \infty$) allows to retrieve continuous hydrodynamical fields and invariants. How is the limit microcanonical measure related to the axi-symmetric Euler equations ? Is it an invariant measure of the axi-symmetric Euler equations ?\\

The answer is positive but not trivial. The very reason why this should be true relies in the existence of a formal Liouville theorem -- \emph{i.e.} an extension of Liouville theorem for infinite dimensional Hamiltonian systems -- for the axi-symmetric Euler equations. An elementary proof concerning the existence of a formal Liouville theorem can be found in \cite{thalabard2013}. \simon{It is a consequence of the explicit Hamiltonian Lie-Poisson structure of the axi-symmetric Euler equations when written in terms of the toroidal and poloidal fields \cite{szeri1988nonlinear,morrison1998hamiltonian}}. The formal Liouville theorem guarantees that the thermodynamic limit taken in a microcanonical ensemble induces an invariant measure of the full axi-symmetric equations. \\

The same issue arises in the simpler framework of the 2D Euler equations. A similar mapping onto a system of vortices that behaves as a mean-field Potts model,  and definition of the microcanonical measure can be found in  \cite{miller1990statistical,ellis2004statistical,bouchet2011statistical}. In \cite{bouchet2010invariant}, it is discussed why the microcanonical  measure is an  invariant measure of the 2D Euler equations. The proof is adaptable to the axi-symmetric  case but goes beyond  the scope of the present paper.\\

It is thus expected that the microcanonical measure of ensembles  of Beltrami spins is an invariant measure of the Euler axi-symmetric equations, therefore  worth of interest. This motivates the present study.

\section{Statistical mechanics of a simplified problem without helical correlations}
\label{sub:microtoy}
\simon{
In the present section, we investigate  a toy measure which corresponds to  a simplified instance of the full (helical) ensemble. In this toy problem,  the total poloidal circulation is the only Helical Casimir that is considered.  The simplification makes the equilibria more easily and pedagogically derived, and provides some intuitive insights about the physics hidden in the Casimir invariants. Besides, the phase diagram that we obtain in this toy,non-helical problem will turn out to be relevant to describe full, helical one. Impatient readers can skip this section and jump directly to section 4, where the main results of the paper are described. 
}
\subsection{Definition of a non-helical toy axi-symmetric microcanonical ensemble}
\label{ssub:microptilde}
\pg
\simon{
For pedagogic reasons, let us here suppose that the microcanonical measure is not constrained by the presence of the whole set of $2K$ Casimirs and kinetic energy but instead only by the Toroidal Areas , the poloidal circulation $X_\tot$ and the total kinetic energy.  This new problem will be much simpler to understand.  The set of $2K+1$ constraints $\mC$ is here replaced by a ``non-helical'' set $\nhC$ of $K+2$ constraints, defined as  
\begin{equation}
\nhC = \{ E, \Ak_{1\le k\le K}, X_\tot=\sum_{k=1}^K X_k\}.
\end{equation}

In this new problem, the correlations between the toroidal and  poloidal degrees of freedom due to the presence of Helical Casimirs are crudely ignored. The only coupling left between  the poloidal and the toroidal fields is a purely thermal one: the only way the fields can  interact with another is by exchanging some of their energy. In order to make this statement more rigorous, we now need to get into some finer details and build explicitly the non-helical microcanonical measure. 
}
\subsubsection{Explicit construction of a non-helical microcanonical measure}
\label{sssub:microptilde_construction}
In order to exhibit a configuration of Beltrami spins $(\sigma_N,\xi_N)$ that satisfies the constraints $\nhC$, it suffices to pick a toroidal configuration $\sigma_N=(\sigma_{N,ij})_{1\leq i,j\leq N}$ with areas $A_k$ and toroidal energy $E_\tor$ together with a poloidal configuration $\xi_N=(\xi_{N,ij})_{1\leq i,j\leq N}$ with a poloidal circulation $X_\tot$ and poloidal energy $E_\pol=E-E_\tor$.
It is therefore natural to introduce the toroidal spaces of configurations $\mG_N^\tor(E,\Ak)$ together with the poloidal spaces of configurations  $\mG_{M,N}^\pol(E,X_\tot)$  as 

\begin{align} 
&\mG_N^\tor(E,\{A_k\})=\{ 
\sigma_N \in \mathfrak{S}_K^{N^2} \mid  \mE_\tor\left(\sigma_N\right)=E \text{ and } \forall k \in \isp{1;K} \, \mathcal{A}_{k}\left[\sigma_{N}\right] = A_{k} \},\\
\text{and }&\mG_{M,N}^\pol(E,X_\tot)=\{ \xi_N \in [-M;M]^{N^2} \mid \mE_\pol \left(\xi_N\right)=E \text{ and }  \mathcal{X} \left[\xi_{N}\right] = X_\tot \}. 
\end{align}

For finite $N$, there is only a finite number of toroidal energies $E_\tor$ for which the space of toroidal configurations $\mG_N^\tor (E,\{A_k\})$ is non empty. The space of bounded Beltrami-spin configurations $\mG_{M,N} (E,\{A_k\})$ is then simply a finite union of disjoint ensembles, that can be formally written as 

\begin{equation}
\mG_{M,N}(E,\{A_k\},X_\tot) = \bigcup_{0\le E_\tor\le E}{ \mG_N^\tor(E_\tor,\{A_k\}) \times \mG_{M,N}^\pol(E-E_\tor,X_\tot)}.
\label{eq:gEAX}
\end{equation}

\paragraph{Definition of the $M,N$-dependent microcanonical measure.} 
\spg
The $M,N$- dependent microcanonical measure $\d\mP_{M,N}$ is defined as the uniform measure on the space of configurations $\mG_{M,N}(E,\{A_k\},X_\tot)$. In order to specify this measure explicitly, we need to define the $M,N$-dependent volume $ \Omega_{M,N}(E,\{A_k\},X_\tot)$ of  $ \mG_{M,N}(E,\{A_k\},X_\tot)$. To do so, we write $\Omega_N^\tor(E,\{A_k\})$ the number of configurations in  $\mG_N^\tor(E,\{A_k\})$ and  $\Omega_N^\pol(E,X_\tot)$ the hypervolume in ${\mathbb R} ^{N^2}$ of $\mG_{M,N}^\pol(E,X_\tot)$, namely  

\begin{align} 
&\Omega_N^\tor(E,\{A_k\})=\!\! \sum_{\sigma_N \in {\mathfrak S}_K^{N^2} } \!\!\mathbf{1}_{\sigma_N \in \mG_N^\tor(E,\{A_k\})}, \\&\text{~and~}
\Omega_{M,N}^{\pol}\left(E,X_\tot \right) = \hspace{-0.3cm}\prod_{(i,j)\in\isp{1;N}^2}\int_{-\infty}^{+\infty} \d\xi_{N,ij}\mathbf{1}_{\xi_N \in \mG_{M,N}^\pol(E,X_\tot)}.
\end{align}
Note that the integral defining the poloidal volume is finite since $\mG_{M,N}^\pol(E,X_\tot)$ is a bounded subset of ${\mathbb R} ^{N^2}$. 
Using equation (\ref{eq:gEAX}), the phase-space volume can then be written as   
\begin{equation} 
\Omega_{M,N} (E,\{A_k\},X_\tot) = \int_0^E \d E_\tor \, \Omega_N^\tor(E_\tor,\{A_k\})\, \Omega_{M,N}^\pol(E-E_\tor,X_\pol).
\label{eq:phasespacevolume}
\end{equation}

\spg

The microcanonical weight $\d\mP_{M,N}(\mC)$ of a configuration $\mC=(\sigma_N,\xi_N)$ lying in the space $\mG_{M,N}(E,\Ak,X_\tot)$ can now be explicitly written as

\begin{equation}
\d\mP_{M,N} ( \mC) =  \dfrac{1}{\Omega_{M,N}\left(E,\Ak,X_\tot \right)}{\prod_{(i,j)\in\isp{1;N}^2}\hspace{-0.3cm} \d\xi_{N,ij}}.
\label{eq:microweight}
\end{equation}

Provided  that ${\mG}$ is a compact subset  of $\mSK^{N^2} \times {\mathbb R}^{N^2}$ it is convenient to  use the shorthand notation  
\begin{equation}
\int_{\mG} \dpmn \equiv \dfrac{1}{\Omega_{M,N}\left(E,\{A_k\},X_\tot \right)} \sum_{\sigma_N\in {\mathfrak S}_K^{N^2}}\, \left({\prod_{(i,j)\in\isp{1;N}^2}\int_{-\infty}^\infty \d\xi_{N,ij} }\right) \mathbf{1}_{\left(\sigma_N,\xi_N \right)\in \mG},
\label{eq:notation} 
\end{equation}
so that the $M,N$ dependent microcanonical average $<>_{M,N}$ of an observable  $\mO $  can now be defined as   

\begin{equation}
\langle \mO \rangle_{M,N} = \int_{\substack{\vspace{0.25cm}\\\mG_{M,N}(E,\{A_k\},X_\tot)}}  \hspace{-2.5cm} \dpmn \mO\left[\sigma_N,\xi_N\right] = \int_0^E \d E_\tor\, \int_{\substack{\vspace{0.25cm}\\ \mG_{N}^\tor(E_\tor,\Ak) \times \mG_{M,N}^\pol(E-E_\tor,X_\tot)}}  \hspace{-4.5cm} \dpmn \mO\left[\sigma_N,\xi_N\right]\label{eq:microptilde_mn}.
\end{equation}

\paragraph{Definition of the limit measures.} 
\spg
It is convenient to use observables to define the limit microcanonical measures.  We define the $M$-dependent microcanonical measure $<>_M$  and the microcanonical measure  $<>$ by letting successively $N\to \infty$ and $M\to \infty$, so that for any observable $\mO$,  $<\mO>_M$ and $<\mO>$ are defined as 
\begin{equation}
 \langle \mO \rangle_{M} = \lim_{N\to\infty} \langle \mO \rangle_{M,N} \text{,~and~}
 \langle \mO \rangle = \lim_{M\to\infty} \langle \mO \rangle_M.
\label{eq:microptilde}
\end{equation}

\subsubsection{Observables of physical interest}
\label{sssection:observables}
Without any further comment about observables and the kind of observables that we will specifically consider, equations (\ref{eq:microptilde_mn}) and (\ref{eq:microptilde}) may appear to be slightly too casual. Let us precise what we mean. In our context, we need to deal both with observables defined for the continuous poloidal and toroidal fields and for their discretized counterparts. 
Given a continuous field $(\sigma,\xi)$,  we consider observables $\mO$ that can be written as $\mO = \int_\mD \dx f^\mO_{\x}(\sigma,\xi)$ where $f^\mO_{\x}$ is a function defined over $\mSK^\mD \times {\mathbb R}^\mD \times \mD$. The discrete counterpart of $\mO$ is then defined as 
\begin{equation}
\mO(\sigma_N,\xi_N) = \dfrac{\vD}{N^2}\sum_{(i,j)\in \isp{1;N}^2} f^\mO_{\xij}(\sigma_N,\xi_N),
\end{equation}
 and the distinction between discrete and continuous observables is made clear from the context.\\
 
To learn about the physics described by the microcanonical measure, a first non trivial functional to consider is the toroidal energy functional $\mE_{tor}$ defined in equation (\ref{eq:discreteenergy}), whose microcanonical average will tell what the balance between the toroidal and poloidal energy for a typical configuration Beltrami spins is. 
In order to specify the  toroidal and poloidal distributions in the thermodynamic limit we will then estimate the microcanonical averages of specific one-point observables, namely 
\begin{equation}
  \mO(\{\sigma\},\{\xi\})=\int_\mD \dx \delta\left({\bf x}-{\bf x_0} \right)\sigma\x^p \xi\x ^k = \mO^\tor(\{\sigma\}) \mO^\pol(\{\xi\})
\label{eq:observables}
\end{equation} with $\mO^\tor(\{\sigma\})= \sigma\xnod^p$ and $\mO^\pol(\{\xi\})= \xi\xnod^k$  defined for any point $\xnod \in \mD$. The microcanonical averages of those observables provide the  moments of the one-point probability distributions and therefore fully specify them.  \footnote{One can observe that one-point moments may be  ill-defined in the discrete case so that their limit may be ill-defined too. One way to deal with this situation is to consider dyadic discretizations, namely choose $N=2^n$. Then for any point $\x$  whose coordinates are dyadic rational numbers, the discrete quantities are non trivially zero when $n$ is large enough. The microcanonical averages can then be extended to any position in  $\mD$ by continuity.} \\

Just as for the 2D Euler equations, and slightly anticipating on the actual computation of the microcanonical measures, we can expect the axisymmetric microcanonical measures to behave as Young measures, that is to say that the toroidal and poloidal distributions at positions $\x$ are expected to be independent from their distributions at position $\xprime\neq \x$. Therefore, specifying the one-point probability distributions will hopefully suffice to completely describe the statistics of the poloidal and of the toroidal field in the thermodynamic limit.

\subsubsection{Specificity of the non-helical toy measure}
\label{ss:ptildespec}
Looking at equation (\ref{eq:microptilde_mn}), it is yet not so clear that our non-helical toy problem  is easier to tackle than the full pronlem,  nor that  the limit measures prescribed by equation (\ref{eq:microptilde}) can be computed. The reason why we should keep hope owes to large deviation theory. Using standard arguments from statistical physics, we argue hereafter that the non-helical problem  can be tackled by defining appropriate poloidal and  toroidal measures that can be studied separately from each other.
\spg
Let us for instance consider the Boltzmann entropies per spin 
\begin{align}
S_N^\tor(E,\{A_k\}) = \dfrac{1}{N^2}\ln &\,\Omega^\tor_N(E,\{A_k\}),~~ S_{M,N}^\pol(E,X_\tot) = \dfrac{1}{N^2}\ln \Omega^\pol_N(E,X_\tot), \label{eq:finite_entrotorpol}
 \\ &\text{~~and~~} S_{M,N}(E,\{A_k\},X_\tot) = \dfrac{1}{N^2}\ln \Omega_N(E,\{A_k\},X_\tot).
\end{align}

As $N\to \infty$, it can be expected that the toroidal entropies $S_N^\tor(E,\{A_k\})$ together with the poloidal entropies $S_{M,N}^\pol(E,X_\tot)$ converge towards a finite limit if they are properly renormalized. If this is the case, then those entropies can be asymptotically expanded as 
\begin{align}
 S_N^\tor(E,\{A_k\}) \underset{N\to \infty}{=} & c^\tor_N(\{A_k\}) + S^\tor(E,\Ak) + o\,(1), \label{eq:entrotor_asymp}\\ 
&\text{~~and~~} S_{M,N}^\pol(E,X_\tot) \underset{N\to \infty}{=} c^\pol_{M,N}(X_\tot) + S_{M}^\pol(E,X_\tot) + o\,(1).
\label{eq:entropol_asymp}
\end{align}

Plugging the entropies into equation (\ref{eq:phasespacevolume}), we get, when $N \to \infty$ 

\begin{equation}
 \Omega_{M,N}(E) = e^{N^2\left( c^\tor_{N}+c^\pol_{M,N}\right)+o(N^2)}\int_0^E \d E_\tor\, \mathrm{e}^{N^2\left(S^\tor(E_\tor) + S_M^\pol(E-E_\tor)\right)}.
\label{eq:volumeentropy}
\end{equation}
For clarity,  we have dropped out  the  $\Ak$ and $X_\tot$ dependence of the different entropies.
Using Laplace's method to approximate integrals, taking logarithm of both sides of equation (\ref{eq:volumeentropy}), dividing by $N^2$, and setting $c_{M,N}\left(\Ak,X_\tot \right) =  c^\tor_{N}\left(\Ak\right)+c^\pol_{M,N}\left(X_\tot \right)$ we obtain 

\begin{align}
 S_{M,N}(E) \underset{N\to\infty}{=} c_{M,N} + & S^\tor(E^\star_M) + S_M^\pol(E-E^{\star}_M) + o(1), \nonumber \\
 &\text{~where~} E^{\star}_M = \underset{x\in[0;E]}{\arg \max}{\{ S^\tor(x) + S_M^\pol(E-x)\}}.
\label{eq:additiveentropy}
\end{align}

A heuristic way of interpreting  equation (\ref{eq:additiveentropy}) is to say that when $N \gg 1$, ``most of'' the  configurations in $\mG_{M,N}(E,\{A_k\},X_\tot)$ have a toroidal energy equal to $E_M^\star$ and a poloidal energy equal to $E-E_M^\star$.

\spg
We can refine the argument, and ask what the typical value of a one-point  observable $\mO = \mO^\tor \mO^\pol$ as described in equation (\ref{eq:observables})  becomes in the thermodynamic limit $N\to \infty$. Let us write the $M,N$ dependent toroidal and poloidal partial microcanonical measures as 

\begin{equation}
 \dpmntor (\sigma_N) = \dfrac{1}{\Omega_N^\tor(E,\Ak)} \text{~and~}  \dpmnpol (\xi_N) = \dfrac{1}{\Omega_{M,N}^\pol(E,X_\tot)} \prod_{(i,j)\in\isp{1;N}^2} \hspace{-0.3cm} \d\xi_{N,ij},
\end{equation}
 
and introduce the shorthand notations 
\begin{align}
&\int_{\mG} \dpmntor \equiv  \dfrac{1}{\Omega_N^\tor(E,\Ak)} \sum_{\sigma_N\in {\mathfrak S}_K^{N^2}}\mathbf{1}_{\sigma_N\in \mG}, \nonumber\\
\text{and } & \int_{\mG} \dpmnpol \equiv \dfrac{1}{\Omega_{M,N}^\pol(E,X_\tot)} \left({\prod_{(i,j)\in\isp{1;N}^2}\int_{-\infty}^\infty \d\xi_{N,ij} }\right) \mathbf{1}_{\xi_N \in \mG}.
\label{eq:notationpartial} 
\end{align}

 Respectively defining the $M,N$ dependent toroidal and poloidal partial microcanonical averages as

\begin{equation}
\langle \mO^\tor \rangle^{\tor,E}_{N} = \int_{\substack{\vspace{0.2cm}\\\mG_{N}^\tor(E,\Ak)}}  \hspace{-1.5cm}\dpmntor \mO^\tor \left[\sigma_N\right]  \text{~~~and~~~} \langle \mO^\pol \rangle^{\pol,E}_{M,N} = \int_{\substack{\vspace{0.2cm}\\\mG_{M,N}^\pol(E,X_\tot)}}  \hspace{-1.5cm}\dpmnpol \mO^\pol \left[\xi_N\right],
\end{equation}

it stems from equation (\ref{eq:microptilde_mn}) that 

\begin{align}
& \langle \mO \rangle_{M,N} =\int_{0}^{E} \d E_\tor \, {\mP}_{M,N} (E_\tor) \, \langle \mO^\tor \rangle^{\tor,E_\tor}_{N} \, \langle \mO^\pol \rangle^{\pol,E-E_\tor}_{M,N}, \label{eq:obs_mn} \\
&\text{~~with~~}{\mP}_{M,N} (E_\tor) = \dfrac{\Omega_N^\tor(E_\tor)\, \Omega_{M,N}^\pol(E-E_\tor)}{\Omega_{M,N}(E)}.
\end{align}

The latter equation means that  the full microcanonical measure $<>_{M,N}$ can be deduced from the knowledge of the partial measures $<>_{N}^{\tor,E}$ and $<>_{M,N}^{\pol,E}$. As $N\to \infty$, the limit measure can be expected to be dominated by one 
of the partial measures,  provided that the limit measures $<>^{\tor,E}$, $<>_{M}^{\pol,E}$ -- defined accordingly to equation (\ref{eq:microptilde}) behave as predicted by equations (\ref{eq:entrotor_asymp}) and (\ref{eq:entropol_asymp}).\\

If for example one considers  an observable $\mO$ that is bounded independently from $N$ ,  then its microcanonical average can be estimated from equation (\ref{eq:obs_mn}) as 
\begin{equation}
\langle \mO \rangle_M = \langle \mO^\tor \rangle^{\tor,E_M^\star} \, \langle \mO^\pol \rangle^{\pol,E-E_M^\star}_{M}.
\end{equation}
  
Thermodynamically stated, this means that the non-helical statistical equilibria  can be interpreted as  thermal equilibria between the toroidal and the poloidal fields. It is therefore relevant to first study  separately the toroidal and the poloidal problem separately from one another. This is what we do in the next three sections.

\subsection{Statistical mechanics of the toroidal field}
\label{ssec:sm_tor_nonhelical}
It is possible to estimate the toroidal entropies $S_N^\tor(E,\Ak)$ for very specific values of the energy using standard statistical mechanics counting methods. We first present those. Then, we show that those specific cases are retrieved with a more general calculation involving the theory of large deviations.

\subsubsection{Traditional counting}
The contribution to the toroidal energy of a toroidal spin $\sigma_{k_0} \in \mSK$  placed at a radial distance  $y =\dfrac{r^2}{2}$ from the center of the cylinder is $\dfrac{\vD {\sigma_{k_0}^2} }{4y N^2}$ . Clearly, the energy is extremal when the $\sigma_k^2$ are fully segregated in $K$ stripes, parallel to the $z$ axis, each of  width $w_k= \dfrac{\left(R^2_{out}-R^2_{in}\right)A_k}{2\vD} + O\left(\dfrac{1}{N}\right) $.   The minimum (resp. maximum) of energy $E_{\min}$ (resp. $E_{\max}$) is  obtained when the levels of $\sigma_k^2$ are sorted increasingly (resp. decreasingly) from the internal cylinder. The number of toroidal configurations that corresponds to each one of those extremal energy states is therefore at most of order $N$. Using definition (\ref{eq:finite_entrotorpol}) and equation (\ref{eq:entrotor_asymp}) , one therefore finds $S^\tor(E_{\min},\Ak) = S^\tor(E_{\max},\Ak)= 0$.
 \spg
Further assuming that $S^\tor(E,\Ak) $ is sufficiently regular on the interval $[E_{\min};E_{\max}]$ , the latter result implies  that there exists an energy value $E^\star\in [E_{\min};E_{\max}]$ for which the  entropy  $\dsp S^\tor(E,\Ak) $ is maximal. The value of $S^\tor(E^\star,\Ak)$ can be estimated by counting the total number of toroidal configurations -- regardless of their toroidal energies  \footnote{We here tacitly work in the case where the $\sigma_k^2$ are all distinct --otherwise  we need to group the levels with the same value of $\sigma_k^2$.}. Indeed,
\begin{align}
 \dfrac{N^2!}{\prod_{k=1}^K N_k!}  =\int_{E_{\min}}^{E_{\max}} \d E \, \Omega_N^\tor (E,\Ak ) & =    \int_{E_{min}}^{E_{max}} \d E \,  e^{N^2 S_N^\tor(E,\Ak)}, \\
& \text{~where~} N_k = \dfrac{N^2A_k}{\vD} \nonumber.
\end{align}
Then, taking the limit $N \to \infty$, using Stirling formula for the l.h.s and estimating the r.h.s with the method of steepest descent, we obtain
\begin{equation}
 S^\tor(E^\star,\Ak) = -\sum_{k=1}^K\dfrac{A_k}{\vD}\ln\dfrac{A_k}{\vD}.
\label{eq:estar_estimate}
\end{equation}

This value corresponds to the levels of $\sigma_k^2$ being completely intertwined.

\subsubsection{Large deviation approach}
\label{par:torsanov}
\pg
We can work out the entropy for any value of the energy by using the more modern framework of large deviation theory.
\spg 
For a given $N$, let us consider the set of random toroidal configurations that can be obtained by randomly and independently assigning on each node of the lattice a level of $\sigma_k$  drawn from a uniform distribution over the discrete set $\mathfrak{S}_K$.  There are $K^{N^{2}}$ such different configurations.  Among those, there exist some that are such that $\forall k \in \isp{1;K}\, \mA_k[\sigma_N] = A_k$ together with $\mE^\tor[\sigma_N] =E$. The number of those configurations is precisely what we have defined as $\Omega_N^\tor(E,\Ak)$.  Can we estimate $\Omega_N^\tor(E,\Ak)$ for  $ N \gg 1$?  The answer is provided by  a large deviation theorem called Sanov theorem -- see for example \cite{ellis1984large,cover1994elements,touchette2009large} for material about this particular theorem and the theory of large deviations.

\spg
Through a coarse-graining, we define the local probability $p_k \x$ that a toroidal spin takes the value $\sigma_k$ in an infinitesimal area $\dx$ around a point $\x$.  With respect to the ensemble of configurations, the functions $(p_1 , ..., p_K )$ define a toroidal macrostate, which satisfies the local normalization constraint:
\begin{equation}
\forall {\textbf x} \, \, \in\mD,  \sum_{k=1}^K p_k\x =1.
\label{eq:localnorm}
\end{equation}
We  denote $\mQ^\tor$  the set of all the toroidal macrostates -- the set of all $p=(p_1 , ..., p_K )$ verifying (\ref{eq:localnorm}). From Sanov theorem, we can compute the number of configurations corresponding to the  macrostate $p=(p_1 , ..., p_K )$. This number is equivalent for large $N$ to the exponential of $N^2$ times the  macrostate entropy 
\begin{equation}
 \mS^\tor[p] = - \dfrac{1}{\vD} \int_\mD\dx \sum_{k=1}^K p_k\x \ln p_k \x.
\label{eq:torentropymacro}
\end{equation}

The toroidal areas $A_k$ occupied by each toroidal patch $\sigma_k$, as well as the toroidal energy constraint, can be expressed as linear constraints on the toroidal macrostates:
\begin{equation}
\forall k \in \isp{1;K}\, \mA_k[p]= \int_\mD\dx p_k\x \text{~and~} \mE_\tor[p]=\int\dx \sum_{k=1}^K p_k\x \dfrac{\sigma_k^2}{4y},
\label{eq:obsmacro}
\end{equation}
where $\mE_\tor[p]$ and $\mA_k[p]$ are the energy and areas of a macrostate $p=(p_1 , ..., p_K )$. As the log of the entropy is proportional to the number of configurations, the most probable toroidal macrostate will maximize the macrostate entropy (\ref{eq:torentropymacro}) with the constraints  $ \dsp \forall k \in \isp{1;K},\, \mA_k[p]=A_k$ and $\dsp \mE_\tor[p]=E$. Moreover, using Laplace method of steepest descent, we can conclude that in the limit of large $N$, the total entropy is equal to the entropy of the most probable macrostate. Therefore,
\begin{align}
 S^\tor(E,\Ak) &= \lim_{N\to\infty}\dfrac{1}{N^2}\ln \Omega_N^\tor(E,\Ak) \\
 &= \sup_{p\in \mQ^\tor} \{\mS^\tor[p] \mid \forall k\in\isp{1;K}\, \mA_k[p]=A_k \text{~and~} \mE_\tor[p]=E\}.
\label{eq:torsanov}
\end{align}

The optimization problem which appears in the r.h.s. of equation (\ref{eq:torsanov}) can be standardly solved with the help of Lagrange multipliers $\alpha_k$ and $\beta_\tor$ to respectively enforce the constraints on the areas $A_k$ and on the energy $E$. The critical points $p^{\star,E}$ of the macrostate entropy for the constraints $E$ and $A_k$ can then be written as 

\begin{align}
 p_k^{\star,E}\x=\dfrac{1}{Z^\star\x}\exp\{\alpha_k-\beta \dfrac{\sigma_k^2}{4y}\} \text{~with~} Z^\star\x = \sum_{k=1}^K \exp\{\alpha_k-\beta \dfrac{\sigma_k^2}{4y}\}.
 \end{align}
 
$\alpha_k$ and $\beta_\tor$  are such that 
\begin{equation}
 \int_\mD\dx \dfrac{\partial \ln Z^\star \x}{\partial\alpha_k} = A_k \text{~and~} -\int_\mD\dx \dfrac{\partial\ln Z^\star \x}{\partial \beta_\tor}=E.  
\label{eq:torlagrange}
\end{equation} 

\spg
Note that if we don't enforce the energy constraint in (\ref{eq:torsanov}), it is easily checked that the maximum value of the macrostate entropy is $\mS^\tor[p^\star]=-\sum_{k=1}^K\dfrac{A_k}{\vD} {\ln \dfrac{A_k}{\vD}}$ obtained for the  macrostate  $p$ defined by  $p_k^\star\x = \dfrac{A_k}{\vD}$. This shows the consistency of our calculation since the latter macrostate can  also be found by setting $\beta_\tor = 0$ in (\ref{eq:torlagrange}). A vanishing $\beta_\tor$ corresponds to the energy constraint $E=E^\star$, so that $ S^\tor(E^\star,\Ak) = - \sum_{k=1}^K\dfrac{A_k}{\vD}\ln\dfrac{A_k}{\vD}$,  and equation (\ref{eq:estar_estimate}) is retrieved. The value of $E^\star$ can be computed from (\ref{eq:obsmacro}) and (\ref{eq:torlagrange}) as $ E^\star = \sum_{k=1}^K \dfrac{A_k\sigma_k^2}{2 \vD} \ln\dfrac{R_{out}}{R_{in}}$.

\begin{figure}[htb]
 \centering
\includegraphics[width=0.6\textwidth]{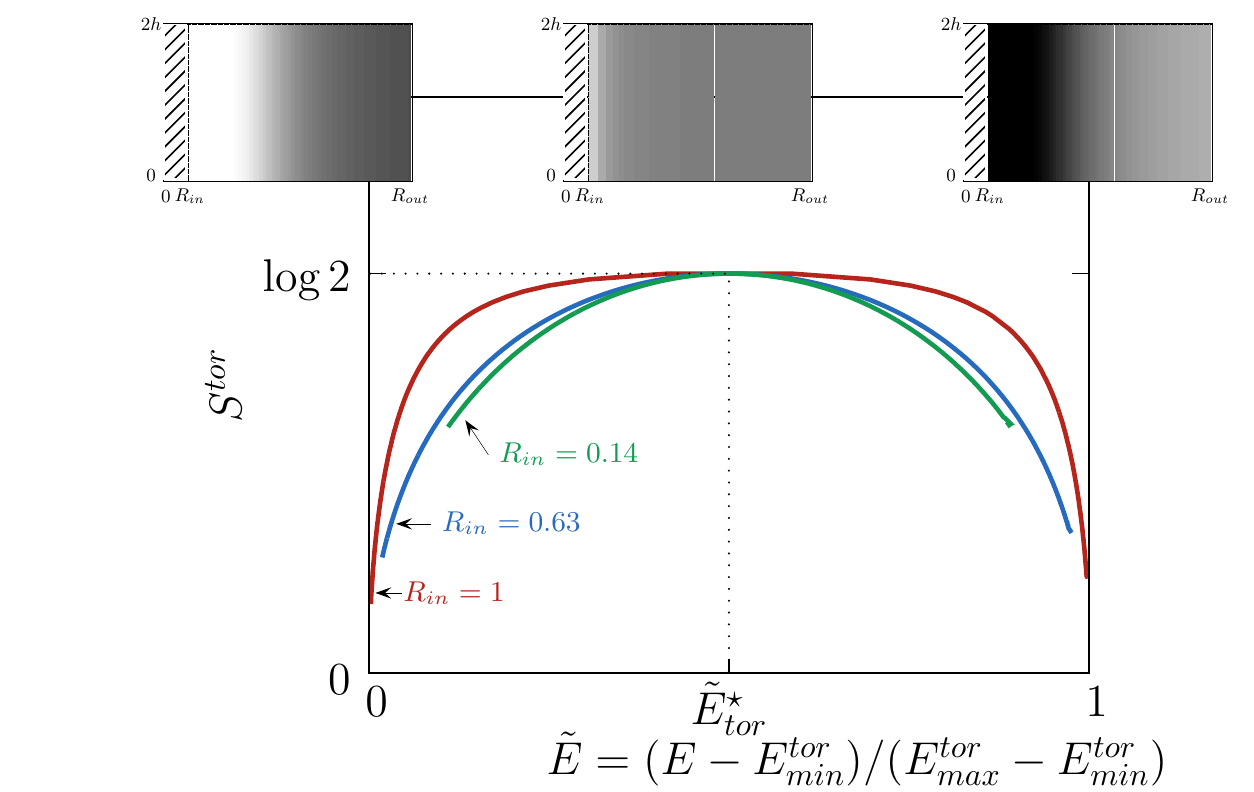} 
\caption{Numerical estimation of the toroidal entropy  for K=2 , $\mathfrak S_2=\{0,1\}$ and $A_0=A_1 = \frac{\mathcal{D}}{2}$. The height of the domain is $2h=1$, its outer radius is $R_{out}=\sqrt 2$ and its inner radius is $R_{in}=0.14$ ,$0.63$ or $1$. Insets show typical toroidal fields $\langle\sigma \x \rangle^{\tor,E}$ for  $R_{in}=0.14$. They correspond to  $E=0.1$, $E=0.5$, and $E=0.9$ from left to right. The grayscale  ranks from  0 (white pixels) to  1  (black pixels).} 
\label{fig:toroidalentropy} 
\end{figure}

\spg
  Equation (\ref{eq:torlagrange}) can also be used to numerically estimate the toroidal entropy for abitrary values of $E$.  Such an estimation is shown on Figure \ref{fig:toroidalentropy} for the specific case where $K=2$, $\mathfrak{S}_2=\{0,1\}$, and $A_0=A_1=\dfrac{\vD}{2}$.

\spg Finally, the microcanonical toroidal moments can be deduced from the critical distribution $p^{\star,E}$ that achieves the maximum macrostate entropy. Those moments read 

\begin{equation}
 \langle  \sigma \x^p \rangle^{\tor,E} = \sum_{k=1}^K  p_k^{\star,E}\x \sigma_k^p.
\label{eq:tormom}
\end{equation}
In the thermodynamic limit, the microcanical measure $<>^{\tor,E} =\lim_{N\to\infty}<>^{\tor,E}_N $ behaves as a product measure, so that equation (\ref{eq:tormom}) completely describes the toroidal micocanonical measure.

 \subsection{Statistical mechanics of the poloidal field}
\label{ssec:sm_pol_nonhelical}
\simon{
The statistical mechanics for the poloidal field is slightly more subtle than for the toroidal field.  It requires two steps: first use a large deviation theorem to compute $<>_M$, then let the cutoff $M$ go to $\infty$.}
\subsubsection{Computation of the $M$-dependent partial measures $<>_M^{\pol,E}$}
 The poloidal energy constraint cannot be exactly expressed as a constraint on the poloidal macrostates. We however argue that Sanov therorem can still be applied because the poloidal degrees of freedom interact through long range interactions, which gives the poloidal problem a mean-field behavior.

\paragraph{}
We consider the set of random poloidal configurations that can be obtained by randomly and independently assigning on each node of the lattice a random value of $\xi$ from the uniform distribution over the interval $[-M,M]$. We then define through a coarse graining  the local probability $\probx$ that a poloidal spin takes a value between $\xi$ and $\xi+\d\xi$   in an infinitesimal area $\dx$ around a point $\x$.  With respect to the ensemble of poloidal configurations, the distributions  $p_M=\{p_M(\xi,\cdot)\}_{\xi \in [-M;M]} $ define a poloidal macrostate.  Each poloidal macrostate satisfies the local normalization constraint :

\begin{equation}
\forall {\textbf x} \, \, \in\mD,  \int_{-M}^M \d \xi \probx =1.
\label{eq:localnormpol}
\end{equation}
We denote $\mQ^\pol$ the sets of all the poloidal macrostates -- the set of all $p_M$ verifying (\ref{eq:localnormpol}).
The number of configurations corresponding to the macrostate $p_M$ is then the exponential of $N^2$ times the poloidal macrostate entropy

\begin{equation}
 \mS_M^\pol[p_M] = -\dfrac{1}{\vD}\int_\mD\dx \int_{-M}^M \d \xi \probx \ln \probx.
\label{eq:polentropymacro}
\end{equation}
The constraint on the total circulation $X_\tot$ can be expressed as a linear constraint on the poloidal macrostates

\begin{equation}
\mX_\tot[p_M]=  \int_\mD \dx \int_{-M}^{M}\d\xi\, \xi \probx .
\end{equation}

\spg
The subtle point arises when dealing with the constraint on the poloidal energy. The energy of a poloidal macrostate is defined as 
\begin{align}
&\mE^\pol[p_M] =\dfrac{1}{2} \int_\mD \dx \psi\x\int_{-M}^{M}\d\xi\, \xi \probx, \label{eq:macroenergy}\\
\text{with ~} &\psi\x = \int_\mD\dx^\prime G({\bf x}, \bf{x}^\prime) \langle \xi\xprime \rangle_M^\pol,\label{eq:selfcty} 
\end{align}
$G({\bf x}, \bf{x}^\prime)$ being the Green function of the operator $-\Delta_\star$ with vanishing boundary conditions on the walls and periodic boundary conditions along the vertical direction. The energy $\mE[\xi_N]$ of a poloidal configuration (\ref{eq:discreteenergy}) is therefore not exactly the energy of the corresponding macrostate (\ref{eq:macroenergy}). In order to deal with this situation, one needs to make the coarse-graining procedure more explicit.
Dividing  the $N\times N$ lattices into $N_b \times N_b$  contiguous blocks each composed of   $n^2=\lfloor N/N_b \rfloor ^2 $ spins, and taking the limit $N \to \infty$ at fixed $N_b$, and then letting  $N_b \to \infty$ , one obtains 
\begin{equation}
\mE^\pol[\xi_N] \underset{\substack{N\to \infty \\ N_b \to \infty }}{=} \mE^\pol[p_M]  + o\left(\dfrac{1}{N_b^2}\right).
\label{eq:meanfield_energy}
\end{equation}

We see that in the continuous limit, the energy of most of the configurations concentrates close to the energy of the macrostate $p_M$ ( see \cite{ellis2000large,potters2012sampling} for a more precise discussion in the context of the 2D Euler equations). This is a consequence of the poloidal degrees of freedom mutually interacting through long range interactions. We can therefore enforce the constraint on the configuration energy as a macrostate constraint.
 
\spg
Following the argumentation yielding to (\ref{eq:torsanov}) in the toroidal case, we conclude that in the limit of large $N$, the total poloidal entropy is equal to the poloidal entropy of the most probable poloidal macrostate which satisfies the constraints. Therefore,
\begin{align}
S^\pol(E,X_\tot)=  \sup_{p_M\in \mQ^\pol} \{\mS^\pol[p] \mid \mX_\tot[p_M]=X_\tot \text{~and~} \mE_\pol[p_M]=E\}.
\label{eq:polsanov}
\end{align}

\paragraph{}
The critical distributions $\pstar$ of the poloidal macrostate entropy can be written in terms of two Lagrange multipliers $\beta\lM_\pol$ and $h\lM$,  respectively related to the constraints on the poloidal energy and on the poloidal circulation as

\begin{align}
 \pstarE = &\dfrac{1}{M \pfunx} \exp\{\left( h\lM-\dfrac{\beta_\pol\lM \psi \x}{2} \right)\xi\}, 
\nonumber  \\
\text{~with ~}  \pfunx  = &\int_{-1}^1 \d\xi \exp\{\left( h\lM-\dfrac{\beta_\pol\lM \psi \x}{2} \right)M\xi\}. \label{eq:reducedpfun}
\end{align}
The Lagrange multipliers $h\lM$ and $\beta_\pol\lM$ are defined through
\begin{align}
 &X_{\tot} = \int_\mD \dx \dfrac{\partial \ln \pfunx }{\partial h \lM} \text{~and~}
 E =-\int_\mD \dx \dfrac{\partial \ln \pfunx }{\partial\beta_\pol\lM}.
\label{eq:lagrangepol}
\end{align}

The moments of the one-point poloidal distribution  can now be estimated from equation (\ref{eq:reducedpfun}) as

\begin{equation}
\forall p \in {\mathbb N},\langle\xi \x^p\rangle_M^{\pol,E} = \int_{-M}^M \d \xi \,\pstar \xi^p = \dfrac{\partial^p \ln \pfunx }{\partial {h\lM}^p}.
\label{eq:momentxi}
\end{equation}

Taking $p =1$ in equation (\ref{eq:momentxi}) and using equation (\ref{eq:selfcty})  yield the $M$-dependent self-consistent mean-field equation 
\begin{equation}
   \dfrac{\partial \ln \pfunx }{\partial {h\lM}} = -\Delta_\star\psi.
\label{eq:mfpol_m}
\end{equation}

We now need to let $M \to \infty$ to describe the microcanonical poloidal measure. \simon{A word of caution may be necessary at this point. For finite $M$, it is possible to estimate the poloidal energy in terms of a macrostate energy as the  correcting term in Equation (\ref{eq:meanfield_energy}) goes to zero when 
$N$ goes to $\infty$. However, the correcting term depends on $M$, which we now want to let go to $\infty$. Therefore, there might be a subtle issue in justifying the rigorous  and uniform decay of the fluctuations of the stream function to zero in the limit $M \to \infty$. In order to make the theory analytically tractable, we will suppose that that such is the case.}

\subsubsection{$M\to \infty$: Computation of the partial limit measures $<>^{\pol,E}$}
\label{sssection:scaling}
We suppose in this section that the energy is non zero.  Otherwise $\psi \equiv 0$ and the equilibrium state is trivial.
\paragraph{Scaling for the Lagrange multipliers.}
\subparagraph{}
\simon{
In order for Equation (\ref{eq:lagrangepol}) to be satisfied whatever the value of $M$, the Lagrange multipliers need to be $M$-dependent. At leading order, the only possible choice is that $\beta_\pol\lM$ and $h\lM$ both scale as $\dfrac{1}{M^2}$, when $M$ goes to $\infty$.\\

 The scaling is crucial to derive the microcanonical equilibria -- whether or not helical.  Let us briefly detail its origin.  It seems  reasonable to assume that $\beta\lM$ and $h\lM$ can be developed in powers of $M$, when $M \to \infty$. Let $\gamma$ be a yet non-prescribed parameter, and let us define  $h^\star$ and $\beta^\star$ as :  

\begin{equation}
 \beta^\star = \lim_{M\to\infty} M^{-\gamma} \beta_\pol\lM \text{~and~} h^\star = \lim_{M\to\infty} M^{-\gamma} h\lM.
\end{equation}

 $h^\star$ and $\beta^\star$ are the first non-vanishing terms in the asymptotic development of $h$ and $\beta$ respectively. They can be interpreted as  ``reduced'' or ``renormalized'' Lagrange Multipliers, associated to the  poloidal circulation constraint and the energy constraint respectively.\\ 

We now consider a fluid element in the vicinity of a point $\xnod$  where the quantity $ f_0^\star = h^\star - \frac{1}{2}\beta^\star\psi\xnod$  is non zero -- this point exists otherwise the stream function $\psi$ would be constant over the domain $\mD$ and the  poloidal energy would be zero. $\psi$ being continuous in the limit $N \to \infty $, we may  assume $\psi\xnod > 0$ on a small volume of fluid $\dxnod$ centered around $\xnod$. To leading order in $M$, this small volume of fluid contributes to the poloidal energy as   
\begin{equation}
E\xnod \dxnod = - \dfrac{\partial \ln Z^{\star}\iM \xnod }{\partial\beta_\pol\lM} \dxnod = \dfrac{M \psi \xnod\dxnod}{2} \dfrac{\int_{-1}^1d\xi \xi e^{f_0^\star M^{\gamma+1}\xi}}{\int_{-1}^1d\xi e^{f_0^\star M^{\gamma+1}\xi}}.
\label{eq:xnodcontri}
\end{equation}
If $\gamma +1 \ge 0$, then $E \xnod \dxnod \to \infty$, and the divergence is exponential when $\gamma>1$. Therefore, $\gamma +1 \le 0$. It stems that   $E \xnod \dxnod \underset{M\to\infty}{\sim} \dfrac{M^{\gamma+2}\psi\xnod f^\star_0}{12} \dxnod$, so that it is finite and non zero only when $\gamma=-2$.\\

Therefore, the correct definition of the reduced Lagrange multipliers, in the case where the poloidal energy is non-vanishing is  

\begin{equation}
  \lim_{M\to\infty}M^2 h \lM=h^\star < +\infty,  \,  \text{~~and~~}  \lim_{M\to\infty}M^2 \beta\lM=\beta^\star < +\infty.
\label{eq:polscaling}
\end{equation}
}


\paragraph{Mean-field equation and infinite temperature}
\subparagraph{}
To describe the microcanonical poloidal measure, we use the scaling (\ref{eq:polscaling}) and let $ M \to \infty$ in Equations (\ref{eq:reducedpfun}) and (\ref{eq:momentxi}). This  yields  
\begin{equation}
 \langle \xi \x \rangle  = -\dfrac{\beta^\star_\pol \psi\x}{6} + \dfrac{h^\star}{3}, \text{~~and~~}   \forall p>1, \, \left| \langle \xi \x^p \rangle \right| =  \infty.  \label{eq:fluctxi}
\end{equation}

The limit mean-field  equation stems from Equation (\ref{eq:fluctxi}) combined with Equation (\ref{eq:mfpol_m}). It reads 
\subparagraph{}
 \begin{equation}
    \Delta_\star \psi = \dfrac{\beta^\star_\pol \psi\x}{6}-\dfrac{h^\star}{3}.
\label{eq:simplifiedclosure}
 \end{equation}

\subparagraph{}
The latter equation is very reminiscent of the equation that describes the low energy equilibria or the strong mixing limit of the 2D Euler equations (see e.g. \cite{chavanis1998classification,bouchet2011statistical}). Standard techniques can be used to solve it. Its solutions are thoroughly determined in Appendix \ref{app:mf_sol}, following a methodology detailed in \cite{chavanis1996classification}. We qualitatively describe those below.
\spg
The differential operator $-\Delta_\star$ is a positive definite operator. We denote by $\phi_{kl}$ and $\kappa_{kl}$  the eigenfunctions and corresponding eigenvalues of $-\Delta_\star$,  such that  $\int_\mD \dx \phi_{kl} \neq 0$. We denote $\phi_{kl}^\prime$ and $\kappa_{kl}^\prime$ the eigenfunctions and corresponding eigenvalues such that $\int_\mD \dx \phi_{kl}^\prime =0$. As shown in Appendix \ref{app:mf_sol}, three kinds of situations can be encountered for a solution $\psi$ of Equation (\ref{eq:simplifiedclosure}).
\begin{itemize} 
\item  If $-\beta^\star/6$ is not one of the eigenvalue $\kappa_{kl}^2$, equation (\ref{eq:simplifiedclosure}) has a unique solution $\psi(\beta^\star,h^\star)$, which is non-zero if $h^\star$ is non zero. If $h^\star \neq 0$, each $\psi(\beta^\star,h^\star)$ can be expressed as a sum of contributions on the modes $\phi_{kl}$ only. This family of solution is continuous for values of $-\beta^\star/6$ between two eigenvalues $\kappa_{kl}^2$, and diverge for $-\beta^\star/6$ close to $\kappa_{kl}^2$. In particular, it is continuous for $-\beta^\star/6=\kappa_{kl}^{\prime,2}$.
\item If $\beta^\star = -6\kappa_{k_0l_0}^{\prime 2}$, $\psi$ is the superposition of the eigenmode $\phi^\prime_{k_0l_0}$ with the solution from the continuum at temperature $\beta^\star = -6\kappa_{k_0l_0}^{\prime 2}$. In this case, $\psi$  is named a ``mixed solution''. 
\item If $\beta^\star = -6\kappa_{k_0l_0}^{2}$,  $\psi$ is proportional to an eigenmode $\phi_{k_0l_0}$. 
\end{itemize}

\paragraph{Entropy and phase diagram.}
\subparagraph{}
All of the solutions described above are critical points for the macrostate entropy. For given $E$ and $X_\tot$ we selected among those critical points those that have the correct $E$ and $X_\tot$. If more than one solution exist, we select the ones that do indeed maximize the macrostate entropy.  The computation of the entropy and the selection of the most probable states is  carried out explicitly in  Appendix \ref{app:KLminimizers}.\\

The type of solutions for which the macrostate entropy is maximal depends on the quantity $\dfrac{X_\tot^2}{2E}$. There exist two threshold values $T_- < T_+$ for this quantity, whose values are here not important but can be found in Appendix \ref{app:KLminimizers}. The value $T_-$ depends on the geometry of the domain. It is close to $T_+$  for thin cylinders $(h \gg R)$ and close to $0$ (but not 0) for wide cylinders  $(h \ll R)$.  We recall that  $\kappa_{01}^2$ is the minimal eigenvalue of the operator $-\Delta_\star$. We denote ${\kappa^{\prime}}^2$ the minimal eigenvalue associated to the eigenfunctions $\phi^\prime$, so that  $\kappa^\prime = \kappa^\prime_{02}$ for wide cylinders and  $\kappa^\prime = \kappa^\prime_{11}$ for thin cylinders.

\begin{figure}[htb]
 \centering
\includegraphics[width=0.49\textwidth]{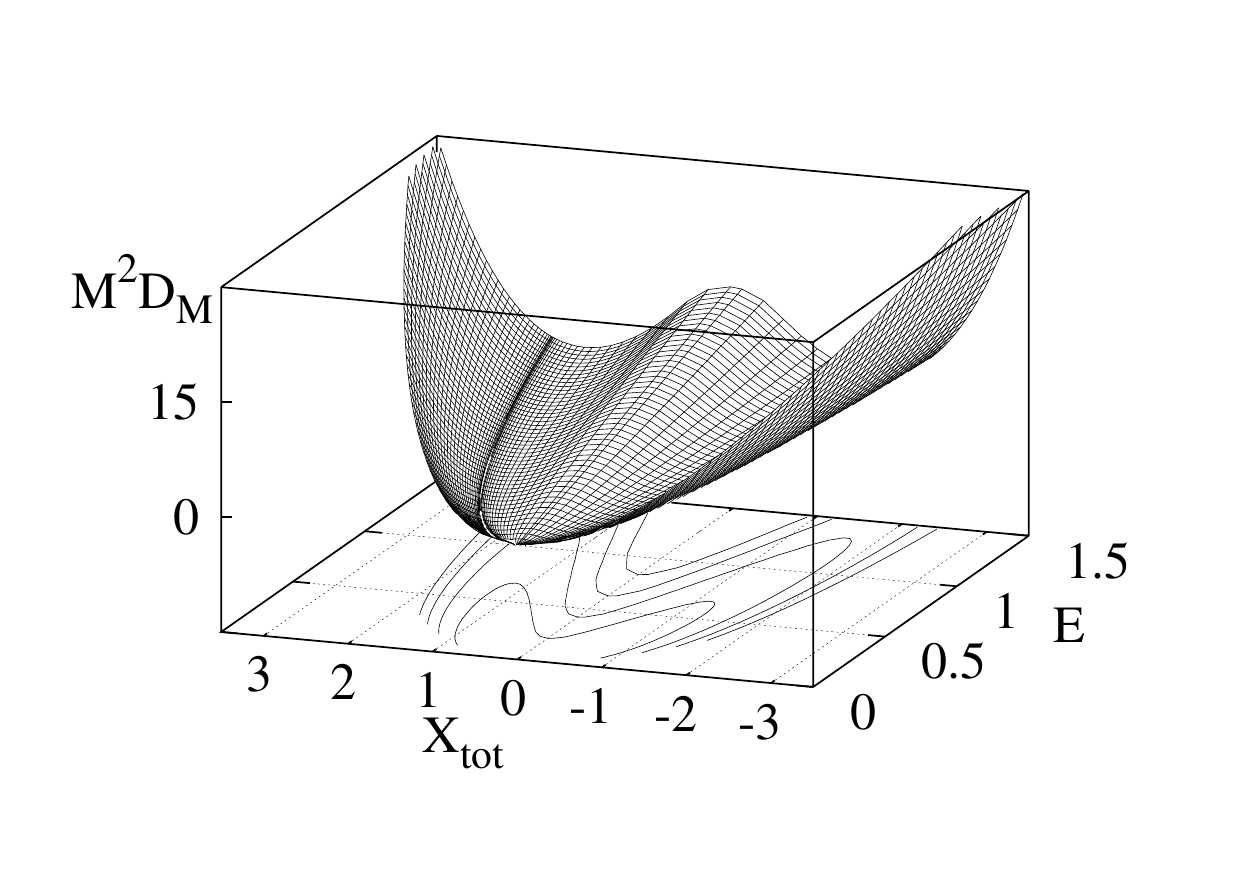} 
\includegraphics[width=0.49\textwidth]{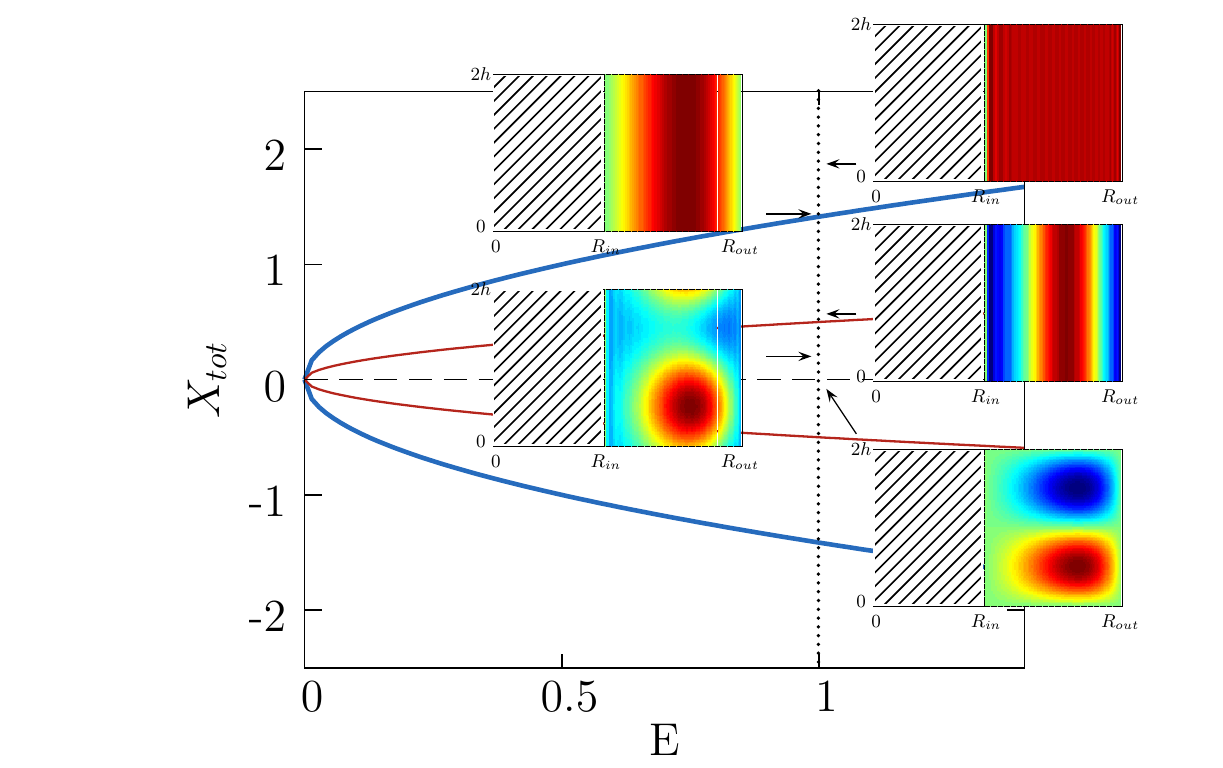} \\
\includegraphics[width=0.49\textwidth]{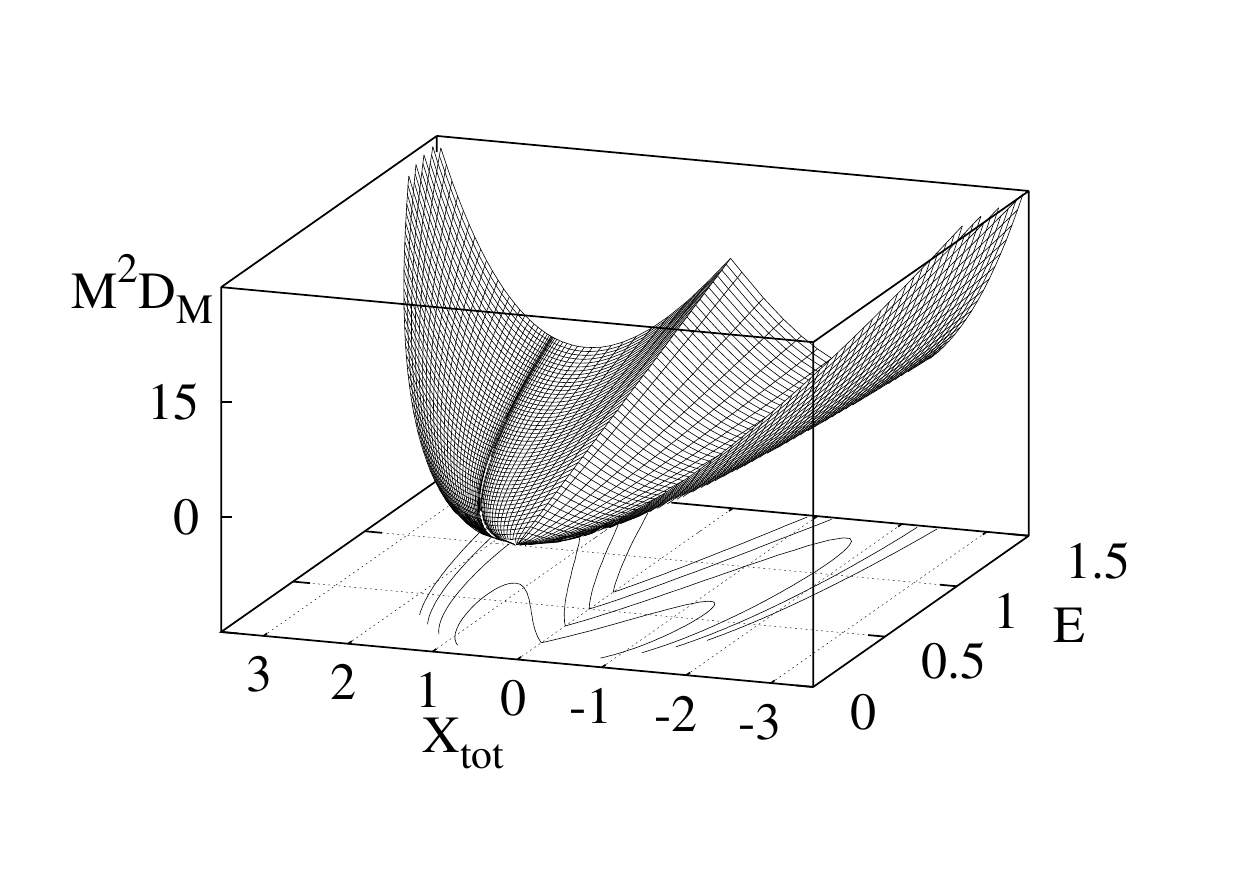} 
\includegraphics[width=0.49\textwidth]{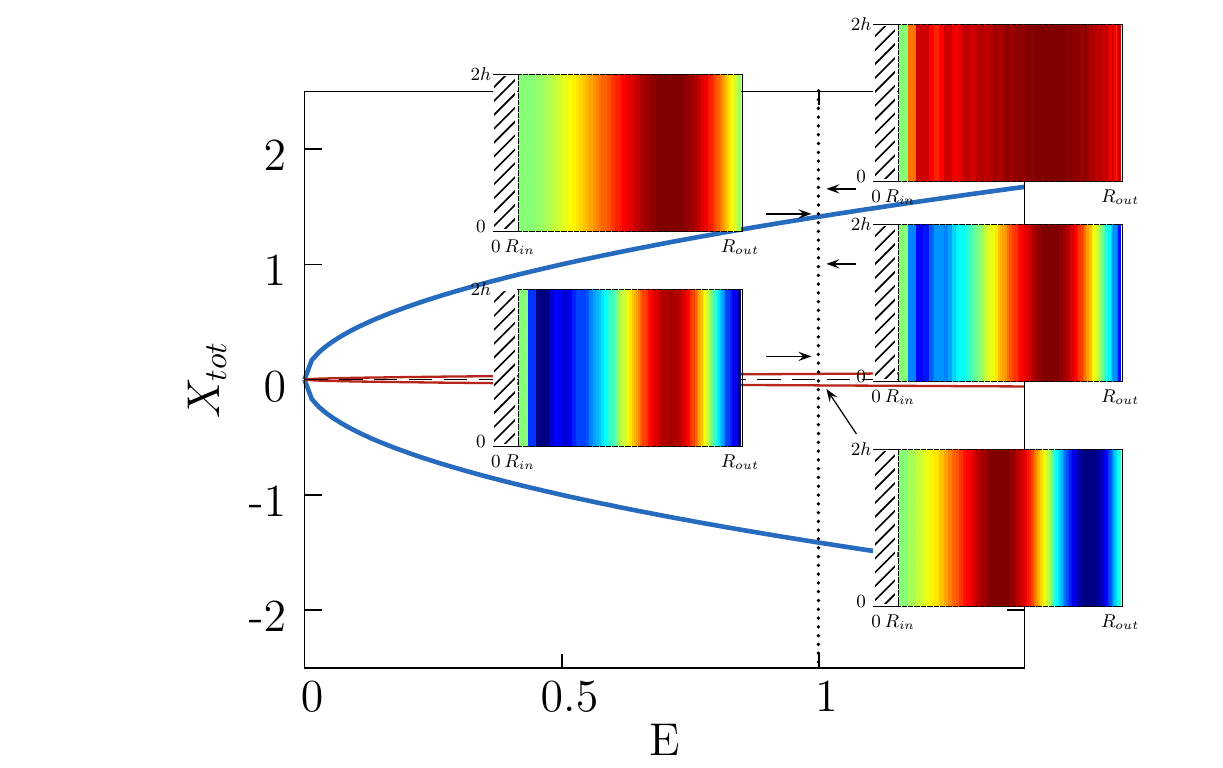} \\
\caption{Left : Minus the poloidal entropy $M^2\mD_M = 2\vD M^2( \ln 2M -\mS_M )$ as a function of the circulation $X_\tot$ and of the poloidal energy $E$. The entropy was numerically estimated for a domain with height $2h=1$, outer radius $R_{out}=\sqrt 2$ and  inner radius $R_{out}=0.63$ (up) and $R_{in}=0.14$ (down) . $X_\tot$ is rescaled by a factor $c_1= \sqrt{\frac{\vD}{32 h}}$  and the entropy by a factor $c_2=\left(\frac{\vD}{2h \pi}\right)^2$ so that the value of $T_+$ is $1$.
Right: The corresponding poloidal phase diagrams. The typical poloidal fields $\langle \xi \x \rangle ^{pol,E}$ are shown $E=1$ and various values of $X_\tot$. Those fields are renormalized by a factor $\sup_\mD \left|\langle \xi\x\rangle ^{\pol,E}\right|$ so that the  colormap ranks from -1 (blue) to 1 (red). With our choice of units the blue parabola has equation $X_\tot^2=2E$. The red parabola separates the solutions from the continuum from the mixed solutions (see text and Appendix \ref{app:KLminimizers} for details).
 } 
\label{fig:phasepol}
\end{figure}

Then:
\begin{itemize}

\item For  $\dfrac{X_\tot^2}{2E} > T_+$, there is only one set of values ($\beta^\star$,$h^\star$) such that the critical points $\psi(\beta^\star,h^\star)$ satisfy the constraints on the energy and on the circulation.  This is a solution from the continuum with  $\beta^\star$ strictly greater than $-6\kappa_{01}^2$. This unique critical point is the entropy maximum. When $\dfrac{X_\tot^2}{2E} \gg T_+$, the typical poloidal field is uniform. As $\dfrac{X_\tot^2}{2E} \to T^+_+$, the typical poloidal field gets organized into a single large-scale vertical jet.
\item For $\dfrac{X_\tot^2}{2E} \in [T_-;T_+] $,  the entropy is maximized for a solution from the continuum. The value of $h^\star$ and $\beta^\star$ are not uniquely determined by the value $\dfrac{X_\tot^2}{2E}$ and the selected solution is the one that corresponds to $|\beta^\star| \le 6 \kappa^\prime$. As $\dfrac{X_\tot^2}{2E} \to T_-^+$, the vertical jet gets thinner.
\item For $\dfrac{X_\tot^2}{2E} \le T_- $, the entropy is maximized by a mixed solution, associated to the eigenvalue $\kappa^\prime$. As $\dfrac{X_\tot^2}{2E} \to 0$, the vertical jet gets transformed into a dipolar flow. The dipoles are vertical for wide cylinders and horizontal for thin cylinders. 
\end{itemize}
Those results and the equilibrium poloidal  fields $\langle\xi \x\rangle^\pol$ are summarized on the phase diagram shown on Figure \ref{fig:phasepol}.
Note, that the entropy of the equilibrium state is 
\begin{equation}
 \mS_M[p_M^{\star,E}] \underset{M\to \infty}{=} \ln 2M + \dfrac{1}{2\vD M^2}\left(\beta^\star E  - h^\star X_\tot \right)  + o\left(\dfrac{1}{M^2}\right),
\label{eq:poloidalentropy}
\end{equation}
where for each value of the energy and of the poloidal circulation, the corresponding values of $\beta^\star$ and $h^\star$ are the ones described above.

\subsection{Statistical mechanics of the simplified problem}
\label{ssub:statmech}
\pg

As explained in Paragraph \ref{ss:ptildespec}, we will now couple the toroidal and the poloidal degrees of freedom in order to solve the non-helical problem and describe the non-helical axi-symmetric measure. The total entropy is then
\begin{equation}
\dsp S_M(E) = \sup_{E_\tor} \{ S_M^\pol(E-E_\tor) + S^\tor(E_\tor) \},
\end{equation}
where $E_\tor$ is the toroidal energy, $E-E_\tor$ the poloidal one. Recall that the toroidal entropy  $S^\tor$ is depicted in Figure \ref{fig:toroidalentropy}, and the poloidal entropy is given by Equation (\ref{eq:poloidalentropy}). The extrema condition leads to the equality of the poloidal and toroidal inverse temperatures 
\begin{equation}
\beta_M^{\pol} = \left. \dfrac{\partial S_M^\pol(E_\pol,X_\tot)}{\partial E_\pol} \right|_{X_\tot} = \beta^{\tor} = \left. \dfrac{\partial S^\tor(E_\tor,\Ak}{\partial E_\tor}) \right|_{\Ak}.We no
\label{eq:temp}
\end{equation}

The fundamental remark is that in the limit $M\rightarrow \infty$, the number of poloidal degrees of freedom scales with $M$. Hence,  the inverse poloidal temperature is equal to zero whenever the poloidal energy is non zero -- see  Equation (\ref{eq:poloidalentropy}) -- and use that $\beta^\star \rightarrow \infty$ for $E_\pol \rightarrow 0$. When the inverse poloidal temperature is zero, so is the inverse toroidal temperature. This prescribes that the toroidal energy  reaches its extremal value $E^\star$ -- see Figure \ref{fig:toroidalentropy}. We are therefore left with two alternatives:
\begin{itemize}
\item $E<E^\star$ then $E_\pol=0$ and  $E_\tor=E$.
\item $E>E^\star$ then $E_\pol=E-E^\star$ and  $E_\tor=E^\star$.
\end{itemize}

\begin{figure}[hbt]
 \centering
\includegraphics[width=0.5\textwidth]{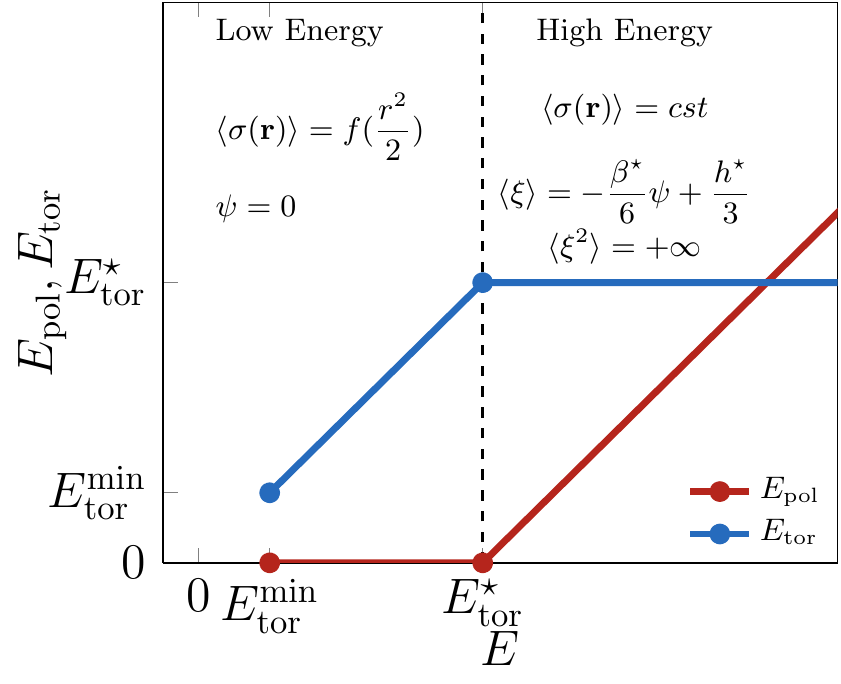} 
\caption{Phase diagram for the non-helical problem. }
\label{fig:phasetilde}
 \end{figure}

\spg The phase diagram corresponding to the non-helical problem is then quite simple, although also quite ``extreme''. It is shown on Figure \ref{fig:phasetilde}, and we can describe the two kinds of equilibria it exhibits. \\

For small energies, (\emph{e.g} $ E < E_\tor^\star$),  there is a large scale organization of the toroidal flow.  In this region, the microcanonical  temperature ${\beta_\tor}^{-1}$ is positive. The smaller $E$ is, the smaller the toroidal temperature is and the less the toroidal energy fluctuates.  As for the poloidal flow, it is vanishing. In the case where $X_\tot$ is non-zero, the limit $E_\pol \to 0$ exists but yields a singular distribution for the poloidal field, since it corresponds to a typical poloidal field having a non-zero momentum while having a vanishing energy.\\

For high energies, (\emph{e.g} $ E > E_\tor^\star$), the equilibria describe toroidal fields that are uniform, the levels of $\mSK$ being completely intertwined. The poloidal fields have infinite fluctuations. This is a consequence of the microcanonical temperature being infinite. When the  poloidal energy is small, typically $ E_\pol \ll \dfrac{1}{2}{X_\tot^2}$, the typical poloidal field is uniform over the domain. For larger poloidal energies, the typical poloidal field gets organized into a single vertical jet ($E_\pol \simeq \dfrac{1}{2}X_\tot^2$) or a large-scale dipole ($E_\pol \gg \dfrac{1}{2}X_\tot^2$).

\section{Statistical mechanics of the full problem}
\simon{
We now consider the full problem, in which the constraints induced by the presence of the Helical Casimirs are no longer ignored.  The construction explicitly carried out in the simplified non-helical case  is easily extended to the general case. A long but straightforward calculation needs to be done to describe the limit microcanonical measure, by letting $N \to \infty$ and $M \to \infty$ subsequently. In the present section, we shall not describe the calculation in full details,  but rather put an emphasis  on  the  main  theoretical results. Quite surprisingly, we find out that the energy phase diagram described in the non-helical case is also relevant for the helical case. In particular, in the high-energy regime, we find out that the correlations play no role in the large scale organization of both fields.  This is quite a striking result which is due to the temperature being infinite whenever the poloidal energy is non vanishing. As a result, the correlations average themselves out at every point of the domain, so that the coarse-grained equilibria only depend on the poloidal  circulation and on the total energy. Some mathematical developments related to the full problem are presented in the next three subsections. The axi-symmetric equilibria are described in (\ref{ssub:phasediagfull}).}

\subsection{Construction of the (helical) axi-symmetric microcanonical measure}
\label{ssub:micropfull}
\pg
Unlike in the previously described non-helical toy problem of Section \ref{sub:microtoy} , the poloidal and the toroidal fields are now coupled not only trough their respective energies, but also through the $K$ partial circulations $\Xk$. In this case, there is no obvious  need to separate the configuration space into a toroidal space and a poloidal space. We therefore cut through this step and directly define the space of bounded Beltrami-spin configurations $\mG_{M,N}(E,\Ak,\Xk)$ together with the  phase space volume $\Omega_{M,N}(E,\Ak,\Xk)$ as

\begin{equation}
\begin{split} 
\mG_{M,N}(E,\Ak,\Xk)&=\left\lbrace\left(\sigma_{N},\xi_{N}\right) \in  \left(\mathfrak{S}_K \times [-M;M]\right) ^{N^2}\right.  \mid  \mE\left(\sigma_{N},\xi_{N}\right) = E \\
  \text{ and } & \left. \forall k \in \isp{1;K}, \, \mathcal{A}_{k}\left[\sigma_{N}\right] = A_{k}\, \text{~and~} \mathcal{X}_{k}\left[\sigma_{N},\xi_N\right] = X_{k} \right\rbrace, \\
\text{and~} \Omega_{M,N}(E,\Ak,\Xk)&= \sum_{\sigma_N \in \mathfrak S_K^{N^2} }\prod_{(i,j)\in \isp{1;N}^2}\int_{-\infty}^{+\infty} \d\xi_{N,ij}  \mathbf{1}_{\left(\sigma_N,\xi_N\right) \in \mG_{M,N}(E,\Ak,\Xk}.
\end{split}
\end{equation}

A straightforward extension of  Equations (\ref{eq:microweight}) and (\ref{eq:microptilde_mn})  is used to define the  microcanonical weight $\d\mP_{M,N}$ of a configuration $\mC=(\sigma_N,\xi_N) \in \mG_{M,N}(E,\Ak,\Xk)$, together with the $M,N$-dependent microcanonical averages $<>_{M,N}$. The microcanonical averages  $<>_M$ and  $<>$ are then defined by letting successively $N\to \infty$ and $M\to \infty$, accordingly to Equation (\ref{eq:microptilde}). 

\subsection{Estimate of $<>_M$}
\spg{}
To describe the limit $N \to \infty$, the central object that we need to investigate is the asymptotic estimate of the phase space volume $\Omega_{M,N}(E,\Ak,\Xk)$. As in the toy problem, we can use a large deviation analysis to relate it to a macrostate entropy. 


\paragraph{}
Randomly and independently assigning on each node of the lattice a random value of $\xi$ from the uniform distribution over the interval $[-M;M]$ together with a random value of $\sigma_k$ drawn from the uniform distribution over $\mSK$, we then define through a coarse- graining procedure the local probability $\probkx$ that a Beltrami spin takes a toroidal value $\sigma_k$ together with a poloidal value between $\xi$ and $\xi+\d\xi$  in an infinitesimal area $\dx$ around a point $\x$.  The distributions  $\dsp p_M=\{p_{k,M}(\xi,\cdot)\}_{\substack{k\in \isp{1;K} \\ \xi \in [-M;M]}} $ define a poloidal macrostate, whose entire set we denote as $\mQ$. The macrostates satisfy the local normalization constraint :

\begin{equation}
\forall {\textbf x} \, \, \in\mD,  \sum_{k=1}^K\int_{-M}^M \d \xi\, \probkx =1.
\label{eq:localnormfull}
\end{equation}

The macrostate entropy is given by

\begin{equation}
 \mS_M[p_M] = -\dfrac{1}{\vD}\int_\mD\dx \sum_{k=1}^K\int_{-M}^M \d \xi \,\probkx \ln \probx/
\label{eq:entropymacro}
\end{equation}

The constraints on the configurations of Beltrami spins can be mapped to constraints on the macrostates through : 
\begin{equation}
\begin{split}
&\mA_k[p_M] = \int_{\mD}\dx \int_{-M}^M \d \xi\, \probkx,  \text{~~} \mX_k[p_M] = \int_{\mD}\dx \int_{-M}^M \d \xi\, \xi \probkx,\\
\text{~and~}&\mE[p_M] = \dfrac{1}{2}\int_{\mD}\dx \sum_{k=1}^K\int_{-M}^M \d \xi\, \{\dfrac{\sigma_k^2}{2y}+\psi\x  \xi\} \probkx.
\label{eq:fullmacroconstraints} 
\end{split}
\end{equation}

The total entropy is then given by the entropy of the most probable poloidal macrostate which satisfies the constraints. Therefore,
\begin{equation}
\begin{split}
S(E,\Ak,\Xk) = \sup_{p_M\in \mQ} \left\lbrace\mS_M[p_M] \mid \forall k\in \isp{1;K}\,  \right. & \mA_k[p_M]=A_k,\,\\ \mX_k[p_M]=X_k 
& \left. \text{~and~} \mE[p_M]=E\right\rbrace.
\end{split}
\label{eq:fullsanov}
\end{equation}
The critical distributions $\pstar$ of the optimization problem (\ref{eq:fullsanov}) can be written using $2K+1$ Lagrange multipliers as

\begin{equation}
\begin{split}
& \pstark = \dfrac{1}{M \pfunx} \exp\{\alpha_k\lM-\dfrac{\beta\lM \sigma_k^2 }{4y}+\left( h_k\lM-\dfrac{\beta\lM \psi \x}{2} \right)\xi\},  \\
\text{with ~}  & \pfunx  = \sum_{k=1}^K \int_{-1}^1 d\xi \exp\{\alpha_k\lM-\dfrac{\beta\lM \sigma_k^2 }{4y}+\left( h_k\lM-\dfrac{\beta\lM \psi \x}{2} \right)M \xi\}, \label{eq:reducedpfun_unplugged}
\end{split}
\end{equation}
where the Lagrange multipliers $\alpha_k\lM$, $h_k\lM$, $\beta\lM$ are determined through

\begin{equation}
\begin{split}
 A_k = \int_\mD \dx \dfrac{\partial \ln \pfunx }{\partial \alpha_k \lM},& \text{~}
 X_k = \int_\mD \dx \dfrac{\partial \ln \pfunx }{\partial h_k \lM},\\
 & \text{~and~} E =-\int_\mD \dx \dfrac{\partial \ln \pfunx }{\partial\beta\lM}.
\label{eq:lagrangepol_unplugged}
\end{split}
\end{equation}
From (\ref{eq:lagrangepol_unplugged}), we can compute the one-point moments as 
\begin{align}
\langle \sigma^p\x \rangle_M = \sum_{k=1}^K\int_{-M}^M \d\xi\,\sigma_k^p\, \pstark \text{~and~} \langle \xi^p\x \rangle_M = \sum_{k=1}^K\int_{-M}^M \d\xi\,\xi^p \pstark.
\label{eq:momentsfull}
\end{align}

In particular, the stream function solves 
\begin{equation}
\Delta_\star \psi\x = - \langle \xi \x \rangle_M = -\sum_{k=1}^K  \dfrac{\partial \ln \pfunx }{\partial h_k\lM}.
\label{eq:mffull}
\end{equation}

Finally, note that the average one-point helicities read : 
\begin{align}
\langle \sigma \x  \xi \x \rangle_M = \sum_{k=1}^K\int_{-M}^M \d\xi\,\xi \sigma_k\, \pstark.
\label{eq:helfull}
\end{align}

\subsection{Estimate of $<>$, and mean-field closure equation}
In order to obtain a microcanonical limit $M\to\infty$ from Equations (\ref{eq:momentsfull}) and (\ref{eq:mffull}), one has to find the correct scaling for the Lagrange multipliers, as derived in the purely poloidal case. We need to consider two cases, depending on whether or not  the poloidal energy $E_\pol$ is vanishing .  
\paragraph{The case $E_\pol \neq 0$.}
With an argument similar to  the one previously exposed in Section \ref{sssection:scaling}, we find out that the correct microcanonical scaling for the Lagrange multipliers is  
\begin{equation}
 \alpha_k = \lim_{M\to\infty} M^0 \alpha_k\lM, ~~ h^\star_k = \lim_{M\to\infty}M^2 h_k\lM,  \,  \text{~~and~~} \beta^\star = \lim_{M\to\infty} M^2 \beta\lM.
\label{eq:fullscaling}
\end{equation}
Using those latter scalings to take the limit $M \to \infty$ in Equations (\ref{eq:mffull}) and (\ref{eq:momentsfull}), one  obtains

\begin{align}
&\forall p \ge 1\, \langle \sigma^p \x \rangle = \overline{\sigma_k^p} \,, \label{eq:toroidalfull} \\
\text{~~together with~~} & \langle \xi\x \rangle = -\dfrac{\beta^\star}{6} \psi\x + \dfrac{1}{3} \overline{h^\star_k}, \text{~~and~~} \forall p \ge 2\,  \left| \langle \xi^p\x \rangle \right| = +\infty,
\label{eq:poloidalfull}
\end{align}
where for any $\{\mO_k\}_{1\le k \le K}$, $\overline{\mO_k}$ is defined by $\dsp \overline{\mO_k} \equiv  \sum_{k=1}^K \dfrac{A_k}{\vD}\mO_k $.
%
%
 %
The closure equation is similar to Equation (\ref{eq:simplifiedclosure})  obtained for the non-helical toy poloidal problem. It reads :  
\begin{equation}
 \Delta_\star \psi=\dfrac{\beta^\star}{6}\psi -\dfrac{1}{3}\overline{h_k^\star}
\label{eq:closure}.
\end{equation}

The one-point helicities are obtained from Equation (\ref{eq:helfull}). They read  
\begin{equation} 
\langle \sigma \x \xi \x \rangle  =  \dfrac{\overline{\sigma_k h_k^\star}}{6} + \langle \sigma \x \rangle \langle \xi \x \rangle.
\end{equation}

Hence, the toroidal and the poloidal fields remain correlated in the limit $M\to \infty$. The first term of the r.h.s  can be interpreted as an extra small-scale contribution to the total helicity. 

\spg
Now, the distributions $p^\star_M$ are critical points of the macrostate entropy (\ref{eq:entropymacro}) but do not necessarily maximize it. We still need to determine which values of $\overline{h_k^\star}$ and $\beta^\star$  actually solve the optimization problem (\ref{eq:momentsfull}), at least for the case under consideration here, that is to say for large values of $M$. 
It turns out, that the asymptotic expansions of the critical values of the macrostate entropy read 
  
\begin{equation}
\begin{split}
 \mS_M [p_M^{\star}]  \underset{M\to\infty}{=}  \ln 2M - \sum_{k=1}^K \dfrac{A_k}{\vD} \ln\dfrac{A_k}{\vD}+ & \dfrac{1}{2\vD M^2}\left(\beta^\star E_\pol - \overline{h_k^\star} X_\tot \right)\\ & + \dfrac{3}{2M^2}  \overline{\left( \dfrac{X_k}{A_k} -  \dfrac{X_\tot}{\vD}\right)^2 }  +  o\left(\dfrac{1}{M^2}\right).
\label{eq:KLfullmacroentropy}
\end{split}
\end{equation}
Some technical details about the derivation can be found in  Appendix \ref{app:macroentropy}. The crucial observation here is that Equation (\ref{eq:KLfullmacroentropy}) compares with the \emph{non-helical} poloidal macrostate entropy given by Equation (\ref{eq:poloidalentropy}). We conclude that the selection of the most probable poloidal state only depends on the value of $E_\pol$ and $X_\tot$. In other words, given a value of $E_\pol$ and $X_\tot$, the most probable  macrostates are the same in the non-helical problem as in the full helical problem, whatever the specific values of the $X_k$ are. 

\paragraph{The case $E_\pol =0$.}
In this case, the stream function $\psi$ is necessarily vanishing. The correct scaling for the Lagrange multipliers is then : 

\begin{equation}
 \alpha_k = \lim_{M\to\infty} M^0 \alpha_k\lM, ~~ h^\star_k = \lim_{M\to\infty}M^2 h_k\lM,  \,  \text{~~and~~} \beta^\star = \lim_{M\to\infty} M^0 \beta\lM.
\label{eq:fullscaling},
\end{equation}

For the toroidal field, such a scaling yields :
\begin{equation}
\langle \sigma^p \x \rangle= \dfrac{\sum_{k=1}^K\sigma_k^p e^{\alpha_k^\star-\beta\sigma_k^2/4y}}{\sum_{k=1}^K e^{\alpha_k^\star-\beta\sigma_k^2/4y}}.
\end{equation}

For the poloidal  field, it yields 
\begin{equation}
\langle \xi \x \rangle= \dfrac{\sum_{k=1}^K h_k^\star e^{\alpha_k^\star-\beta\sigma_k^2/4y}}{3\sum_{k=1}^K e^{\alpha_k^\star-\beta\sigma_k^2/4y}} \text{,~and~}
\langle \xi^p \x \rangle = +\infty \text{~ for  $p >1$ }.
\end{equation}.

Just like in the toroidal problem which was treated in the non-helical case described in Section \ref{ssec:sm_tor_nonhelical},  the Lagrange multipliers $\alpha_k^\star$ and $\beta^\star$ are then uniquely determined by inverting the system made of the $K+1$ equations 
\begin{equation}
\begin{split}
&E = \int_\mD \dfrac{\sum_{k=1}^K (\sigma_k^2/4y) e^{\alpha_k -\beta^\star \sigma_k^2/4y}}{\sum_{k=1}^K e^{\alpha_k -\beta^\star \sigma_k^2/4y}}, \\
\text{~and~} &\dfrac{A_k}{\vD} = \int_\mD \dfrac{e^{\alpha_k -\beta^\star \sigma_k^2/4y}}{\sum_{k=1}^K e^{\alpha_k -\beta^\star \sigma_k^2/4y}} \text{~ for all $1 \le k \le K$}.
\end{split}
\label{eq:lagrangelow}
\end{equation}
It is not difficult to check that the reduced Lagrange multipliers $h_k^\star$  satisfy $h_k^\star = 3 \dfrac{X_k\vD}{A_k}$.\\
 
 Therefore, in the case where the poloidal energy is vanishing, the helical correlations do not affect the typical toroidal states: the toroidal equilibria are exactly those described in Section \ref{ssec:sm_tor_nonhelical} and depicted in a simplified two-level case on Figure \ref{fig:toroidalentropy}. The poloidal field is however ``enslaved'' to the toroidal field. It does not contribute to the total energy. 
 

\subsection{Phase diagram of the full problem}
\label{ssub:phasediagfull}
\paragraph{}
In the last section, we have obtained that in the case where the poloidal energy is non-vanishing that the toroidal levels  $\sigma_k$ are completely mixed -- Equation (\ref{eq:toroidalfull}). As a consequence, $E_\tor=E^\star$. We thus deduce the same alternative as in the reduced problem:
\begin{itemize}
 \item If $E \ge E_\tor^\star$, then  $E_\tor = E_\tor^\star$ and $E_\pol = E-E_\tor^\star$.
 \item If $E < E_\tor^\star$, then $E_\tor = E$ and $E_\pol = 0$. 
\end{itemize}
$E_\tor^\star$ is computed from Equation (\ref{eq:toroidalfull}) as $E_\tor^\star = \sum_{k=1}^K \dfrac{A_k\sigma_k^2}{2 \vD} \ln\dfrac{R_{out}}{R_{in}}$, just as in the non-helical case.  Therefore, the phase diagram describing the splitting of the total kinetic energy  between the toroidal and the poloidal degrees of freedom is exactly the same as the one described in the  simplified problem of Section \ref{sub:microtoy}. It is therefore shown on Figure \ref{fig:phasetilde}. It displays a high energy ($E\ge E^\star$) and a low energy regime ($E<E^\star$). In each of those energy regime, the axi-symmetric equilibria are very much akin to the non-helical  equilibria  described in Section \ref{ssub:statmech}, with just a small alteration for the typical poloidal field in the low energy regime. To make this result stand more  clearly, we summarize below the characteristics of both regimes.  \\
 
\paragraph{In the low energy regime $(E < E^\star)$,} the typical fields are characterized through 
\begin{equation}
\begin{split}
&\langle \sigma \x \rangle= \dfrac{\sum_{k=1}^K\sigma_k e^{\alpha_k^\star-\beta\sigma_k^2/4y}}{\sum_{k=1}^K e^{\alpha_k^\star-\beta\sigma_k^2/4y}}, \text{~} \langle \xi \x \rangle  =  \dfrac{\sum_{k=1}^K h_k^\star e^{\alpha_k^\star-\beta\sigma_k^2/4y}}{\sum_{k=1}^K e^{\alpha_k^\star-\beta\sigma_k^2/4y}} (y),\\
& \text{~ and } \psi = 0.
\end{split}
\label{eq:lowtyp}
\end{equation}

The Lagrange multipliers are determined through Equation (\ref{eq:lagrangelow}). The poloidal fluctuations  are infinite.
Qualitatively, the flow (poloidal and toroidal) is stratified along the radial direction. In the limit of a very low energy ($E \gtrsim E_\tor^{\min}$) the toroidal patches are completely segregated, and sorted by increasing toroidal values from the inner to the outer wall. When the energy gets close to $E_\tor^\star$, it becomes uniform -- see Figure \ref{fig:toroidalentropy}.

\paragraph{In the high energy regime $(E \ge E^\star)$,}
the typical fields are characterized through 
\begin{equation}
\begin{split}
&\langle \sigma \x \rangle= \sum_{k=1}^K \dfrac{A_k}{\vD}\sigma_k, \text{~} \langle \xi \x \rangle  = -\dfrac{\beta^\star}{6}\psi \x + \dfrac{1}{3}\sum_{k=1}^K\dfrac{A_k}{\vD} h_k^\star ,\\
& \text{~ with } \Delta_\star \psi = \dfrac{\beta^\star}{6}\psi \x - \dfrac{1}{3}\sum_{k=1}^K\dfrac{A_k}{\vD} h_k^\star.
\end{split}
\label{eq:hightyp}
\end{equation}

The Lagrange multipliers $\beta^\star$ and $h_k^\star$ can be completely determined -- see Appendix \ref{app:macroentropy}.
The poloidal energy is prescribed as $E_\pol= E-E_\tor^\star$. Qualitatively, the toroidal field is uniform. This corresponds to the toroidal patches being completely intertwined, regardless of their position in the domain $\mD$. The poloidal field exhibits infinitely large fluctuations around a large scale organization. The latter is completely prescribed by the values of the poloidal energy and of the poloidal circulation and does not depend on the specific choice of the partial poloidal circulations $X_k$.   \\

For prescribed values of the constraints, the entropy of the full problem  as given by Equation (\ref{eq:KLfullmacroentropy}) matches the non-helical poloidal entropy (\ref{eq:poloidalentropy}) up to some constants terms. Therefore, the large scale organization of the poloidal field is exactly the one depicted on Figure \ref{fig:phasepol}.

\subsection{Further Comments}
\subsubsection{Stationarity and formal stability of the equilibria}
\paragraph{}
We can observe that the axi-symmetric statistical equilibria described in the previous Section \ref{ssub:phasediagfull} describe average fields which are  stationary states of the Euler axi-symmetric equations (\ref{eq:axi}). In the low energy regime, this is due to the stream function $\psi$ being vanishing and to the typical toroidal field being a function of the radial coordinate only.  In the high energy regime, this is due to the typical toroidal field being constant, and to the poloidal field being a function of the stream function $\psi$. Note that this is in itself a result, and not an input of the theory. \\

Besides, we can also note that not only are those typical fields stationary, they are also formally stable with respect to any axi-symmetric perturbation. For infinite dimensional systems, formal stability is a pre-requisite for non-linear stability \cite{holm1985nonlinear}.  In the case of axi-symmetric flows, a sufficient criterion for formal stability based on the general Energy-Casimir method can be found in \cite[Eq 3.15]{szeri1988nonlinear}. With the notation at use in the present paper, and with an ``e'' subscript to denote an axi-symmetric stationary solution,  this criterion reads 
\begin{equation}
\dfrac{\partial \xi_e}{\partial \sigma_e}\dfrac{\d \psi_e}{\d \sigma_e}+ \dfrac{\sigma_e}{2y^2}\dfrac{\partial y}{\partial \sigma_e}- \dfrac{1}{-\Delta_\star^{-1}}\left(\dfrac{\d\psi_e}{\d\sigma_e}\right) ^2\ge 0. 
\label{eq:formalstability}
\end{equation}
 The notation $1/(-\Delta_\star^{-1})$ can be liberally replaced by any $1/\kappa_i^2$ with $\kappa_i^2$ either one of the eigenvalue of $-\Delta_\star^{-1}$, which are real and non-negative -- see Appendix \ref{app:mf_sol}.
As noticed by Szeri and Holmes, ``the inequality cannot be expected to hold in general, for the simple reason that 
the eigenvalues of the operator [$1/\Delta_\star^{-1}$] have no upper bound''.  However, the criterion is fulfilled for the very limited set of equilibria obtained from our statistical mechanics approach. In the low energy regime, only the term $\dfrac{\langle \sigma \rangle }{2y^2}\dfrac{\partial y}{\partial \langle \sigma \rangle}$ is non-vanishing. It is however positive, as the stratification causes the values of $\langle \sigma \rangle $ to increase from the inner to the outer cylinder. Hence the criterion is fulfilled. 
In the high energy regime, every term involved in Equation (\ref{eq:formalstability}) vanishes. Therefore, the stability criterion is also -- trivially -- fulfilled.

\subsubsection{Link to previous work}
The axi-symmetric equilibria (\ref{eq:hightyp}) and (\ref{eq:hightyp}) which we obtained in the present paper substantially differ from the ones described in previous works about the statistical mechanics of axi-symmetric swirling flows. We can note that an attempt to bound the poloidal fluctuations with an extraneous cutoff can be found in \cite[Appendix E]{leprovost2006dynamics}. In this appendix, a set of canonical equilibria are derived and the authors assume that  a physical interpretation can be given to the extraneous cutoff. Those canonical equilibria are however ``dramatic'' : they depend exponentially on the extraneous cutoff. The authors note that the average fields which are described by this statistical mechanics approach are \emph{not} steady solutions of the axi-symmetric Euler equations. \\

For this reason, \cite{leprovost2006dynamics,
naso2010statistical} rather prefer to work out  the statistical mechanics of the axi-symmetric Euler equations by analogy with the 2D Euler equations, setingthe poloidal fluctuations $0$, and considering a toroidal mixing subject to a ``robust'' set of three constraints, namely the energy, the helicity and the toroidal momentum.  
In \cite[Eq (36-37)]{naso2010statistical}, it is found that the typical fields correspond to large scale Beltrami flows, such that $\langle \sigma \x \rangle =  B \psi \x $ and $\langle \xi \x\rangle  = B \langle \sigma \rangle/ 2y +C$, where $B$ and $C$ are  related to the Lagrange multipliers associated to the constraints of energy, helicity and angular momentum. From a physical point of view, and as far as the axi-symmetric Euler equations are concerned, those equilibria have in a sense two ``drawbacks'' :
i) they predict a multi-stability of solutions and do not predict the emergence of large scale structure as maximal entropy structures and 
ii) they predict that the average fields are  steady states  of the Euler axi-symmetric equations, yet of an unstable kind. More explicitly, for the Beltrami flows just described, the presence of a large scale helicity creates a dependence between the typical toroidal field and the stream function. This makes   the term $\left(\dfrac{\d \psi}{\d \sigma}\right)^2$ in the criterion (\ref{eq:formalstability}) be non vanishing and hereby prevents the steady states from being stable.
\\

In our statistical approach, both of the issues have been fixed, although their outcome was not \emph{a priori} known.  The main ideas were to consider infinitely large poloidal fluctuations, and to work exclusively in the microcanonical ensemble so as to find out a good scaling for the Lagrange multipliers at stake. Besides, we managed to take into account all the invariants. The price to pay is that the equilibria that we get within our approach are in a sense more extreme and more limited than the ones previously found. They are however more natural. 

\section{Discussion}
\paragraph{Some additional technical comments.}
\subparagraph{}
It was not obvious from the beginning that the construction of microcanonical measures \emph{à la} Robert-Miller-Sommeria for the axi-symmetric Euler equations could be carried out extensively, nor that it would yield non trivial insights to understand the physics of axi-symmetric flows.  What can be considered as the key point here is the accurate renormalization of the inverse temperature and associated Lagrange multipliers with respect to the phase space volume. This allowed us to build an asymptotic limit consistent with the physical constraints and prevented us from encountering an avatar of the Jeans paradox. The renormalization was not carried out in the previous works concerning axi-symmetric equilibria. Here, it is crucial in order to take into account the invariants related to the poloidal degrees of freedom that live in an infinite phase space.\\

Other choices could have been made to renormalize the phase space. Instead of a cutoff $M$, it is also possible to make the  divergent  integrals converge by integrating over the $\nu$ dependent measures $e^{-\nu\xi^2} \d\xi$ -- rather than over the $M$ dependent measures ${\mathbf 1}_{[-M;M]} \d\xi$ -- . This is tantamount to restricting the set of macrostates on which the suprema of the entropy are taken, to those whose poloidal fluctuations are bounded.  To work out the microcanonical limit, one then needs to introduce some $\nu$-dependent Lagrange multipliers $\beta^{\nu}=\nu\beta^\star$, $h^{\nu}=\nu h^\star$ and  let $\nu \to 0$ subsequently. The limit measures obtained with the latter renormalization are completely consistent with the ones we described in this paper. They are also in a sense more general as they allow to retrieve the previously found Beltrami states by considering the other limit $\nu \to 0$. \\

Note also that in order to carry out our analysis, we have restricted ourselves to the case where the inner cylinder has a non-vanishing radius $R_{in}$, so that we worked in the framework of a ``Tayor-Couette geometry''. It is yet not so clear how to extend the analysis to the limit case  $R_{in} \to 0$, which can be thought of as a  ``von K\'{a}rm\'{a}n geometry''.  The problem comes from the blow up of the equilibrium toroidal energy $ E_\tor^\star = \sum_{k=1}^K \dfrac{A_k\sigma_k^2}{2 \vD} \ln\dfrac{R_{out}}{R_{in}}$ if we simply let $R_{in} \to 0$. \footnote{One naive way to cope with this issue and obtain a specific class of equilibria for the von K\'{a}rm\'{a}n geometry is to renormalize each toroidal level $\sigma_k^2$ in $\mSK$ as $\sigma_k^2 \to \dfrac{\sigma_k^2}{\ln\dfrac{R_{out}}{R_{in}}}$. Another possibility is to impose a local smoothing condition near the center of the cylinder that could be enforced  at the level of the macrostates. It would suffice for instance to prescribe $<\sigma\x>_M \underset{r\to 0}= O(r^\epsilon)$ with $\epsilon$ being non negative in order to avoid a blow up of the equilibrium toroidal energy. A third alternative is to rule out the existence of infinite temperature states in this geometry. }

\paragraph{Physical insights about axi-symmetric turbulence}
\subparagraph{}
The physics described by the micrononical measure is interesting. Let us first comment about the role of the invariants.
We may have built a measure by taking into account every kind of inviscid invariant of the axi-symmetric Euler equations, it turns out that most of the physics comes from a reduced set of invariants, namely the energy, the toroidal Casimirs and the total circulation.
In particular, our result shows that  the helicity -- which relates to the correlation between the toroidal and the poloidal degrees of freedom  --  plays   no role in the description of large scale structure at the level of the macrostates when the energy is high enough. This is consistent with the traditional picture of a downward helicity cascade in $3D$ turbulence. This may also explain why previous attempts to find axi-symmetric equilibria by neglecting the fluctuations of the poloidal field while keeping a constraint on the helicity would only lead to unstable equilibria, likely to be destabilized by small-scaled perturbations.  
\spg
The axi-symmetric equilibria are very different from those obtained in the 2D case. In the low temperature, low energy regime, the large scale stripes come from the interaction of the toroidal degrees of freedom with the position field --  the interaction being inhomogeneous and invariant with respect to vertical translations. As for the infinite temperature, high energy regime, the Toroidal Casimirs play no role in it. The linear relationship between the poloidal field and the stream function may be seen as the axi-symmetric analogue of the low energy limit of the $\sinh$-Poisson relation in 2D turbulence. Yet, the infinite fluctuations related to the poloidal field may be heuristically interpreted as a very 3D turbulent feature and may be related to the tendency of vortices to leak towards the smallest scales available in 3D turbulence. Therefore, neither regimes have strict analogues in 2D. 

\paragraph{Some perspectives.}
\subparagraph{Extensions to closely related flows.}
Let us mention the close analogy between axi-symmetric flows and other flows of geophysical and astrophysical interests such as two-dimensional stratified flows  in the Boussinesq approximation (Boussinesq flows) \cite{szeri1988nonlinear,abarbanel1986nonlinear} and two-dimensional  magnetohydrodynamics (2D MHD). In the former case, it almost suffices to replace the word ``poloidal'' by the word ``vorticity'' and the word ``toroidal'' by the word ``density '' in the present paper to obtain \emph{ mutatis mutandi}  a statistical theory for ideal Boussinesq flows. The case of 2D MHD is slightly more subtle. The  Casimir invariants of ideal 2D MHD are similar to the axi-symmetric Casimir invariants but the energies slightly differ. It would therefore be very interesting to generalize the method described in the present paper to the 2D MHD case, which is more documented than the axi-symmetric case, and for which inviscid statistical theories have recently been reinvestigated \cite{weichman2012long}. 

\subparagraph{Are microcanical measures relevant for real turbulence ?}
It is finally tempting to ask whether some of the axi-symmetric equilibrium features can be recognized in real turbulent experiments. Examples of a turbulent flows likely to be modeled  by the axi-symmetric Navier-Stokes are von K\'{a}rm\'{a}n turbulence \cite{herbert2012dual,saint2013forcing} or Taylor-Couette turbulence \cite{smith1982turbulent,Dong2007}.  There however exist many caveats concerning a thorough investigation of the link between  axi-symmetric ideal measures and turbulent experiments, examples of which include  requirements on a ``separation of scales'', the relevance of fragile invariants in the presence of forcing and dissipation, the intrinsic ``3Dness'' of a turbulent experiment. We therefore 
postpone the discussion to a forthcoming paper.

\paragraph{Acknowledgements.}
We thank J. Barré, P-H. Chavanis, B. Turkington, A. Venaille and an anonymous referee for their careful proof reading and their useful comments that helped improve the presentation of the arguments discussed in the  present paper.

\newpage
\appendix

\section{Solutions of the mean-field equation}
\label{app:mf_sol}

\paragraph{}
We show here how  to solve the  closure equations (\ref{eq:simplifiedclosure}) and (\ref{eq:closure}) in terms of the eigenmodes of the operator $\Delta^\star$,  for fields that are $2h$-periodic along the $z$ direction and are vanishing on both the inner and the outer cylinders. Recall that those equations both read 
\begin{equation}
 \Delta_\star\psi = \dfrac{\beta^\star}{6}\psi - \dfrac{h^\star}{3} \text{~~with~~ } \Delta_\star = \dfrac{1}{2y}\partial_{zz} + \partial_{yy}.
\label{eq:closure_reminder}
\end{equation}

\subsection{Explicit computation of the eigenmodes of the operator $\Delta^\star$}
\paragraph{}
The eigenmodes of $\Delta_\star$  are solutions to the  eigenvalue problem $\Delta_\star \phi_\kappa = -\kappa^2 \phi_\kappa$ -- with the prescribed boundary conditions. Let $\phi_\kappa$ be such an eigenmode. We can Fourier decompose $\phi_K$ and write  $\dsp \phi_K(y,z)= \sum_{k\in \mathbb Z} f_k(y) \exp{\dfrac{i k \pi z}{h} }$.  $\phi_K$ is a solution to the eigenvalue probleme \emph{iff} each one of the functions $f_k$ satisfies 
\begin{equation}
\begin{split}
 &f_k^\pprime(y) + \left(\kappa^2-\dfrac{k^2 \pi^2}{2h^2y} \right)f_k(y)=0,\\
\text{or equivalently~}  &\tf_k^\pprime(\ty) + \left(1-\dfrac{k^2 \pi^2}{2h^2 \kappa \ty } \right)\tf_k(\ty)=0
 \text{~putting~}  \ty =\kappa y \text{~and~} \tf_k(\ty) = f_k(y).
\label{eq:coulombdim}
\end{split}
\end{equation}
The latter equation is known as a ``Coulomb Wave equation'' \cite{abramowitz1965handbook}.

\subparagraph{If $k=0$,} then $ \dsp f_0(y) = A \sin  \kappa \left(  y-Y_{in}\right) + B \cos \kappa\left(  y-Y_{in}\right)$.
 $f_0(Y_{in})=0$ gives $B=0$. $f_0(Y_{out})=0$ gives $\kappa = \kappa_{0l}=\dfrac{l \pi}{Y_{out}-Y_{in}} $. 
For  each value of $l\ge 0$, we write $\phi_{0l}=\dfrac{\sin \left[\kappa_{0l} \left( y-Y_{in} \right) \right]}{\sqrt{h(Y_{out}-Y_{in})}}$.  $\phi_{0l}$ is an eigenmode of $\Delta^\star$, such that  $\Delta_\star \phi_{0l} = -\kappa^2_{0l} \phi_{0l}$. The normalization factor is chosen so that $\dsp \int_{Y_{in}}^{Y_{out}} dy\int_0^{2h} dz  \phi_{0l}^2 =1$. 

 \subparagraph{If $k \neq 0$, } $\tf_k(\ty) =  C_1 F_0\left(\eta_k,\ty \right) + C_2 G_0\left(\eta_k,\ty \right) $ where $F_0$ and $G_0$ are respectively the regular and singular Coulomb Wave functions associated to the parameter $\eta_k = \dfrac{k^2\pi^2}{4h^2 \kappa}$. The non trivial solutions are determined using the vanishing boundary condition for $\psi$ on the walls. For each value of $k$,  the horizontal eigen modes correspond to the values  $\kappa_{kl}$ for wich the quantity 

\begin{equation}
W(\kappa)=\begin{vmatrix}
F_0\left(\dfrac{k^2\pi^2}{4h^2 \kappa},\kappa Y_{in}\right) & G_0\left(\dfrac{k^2\pi^2}{4h^2 \kappa},\kappa Y_{in}\right)\\
F_0\left(\dfrac{k^2\pi^2}{4h^2 \kappa},\kappa Y_{out}\right) & G_0\left(\dfrac{k^2\pi^2}{4h^2 \kappa},\kappa Y_{out}\right)
\end{vmatrix}
\text{~~is vanishing}.  
\end{equation}

Each mode $\kappa_{kl}$ is therefore related to two eigenmodes  
$\phi_{kl}^\pm = A_{kl}\exp\left({\pm i\dfrac{k\pi z}{h}}\right) f_{k}(\kappa_{kl}y) $,  such that $\Delta_\star \phi_{kl} = -\kappa_{kl}^2 \phi_{kl}$. The normalization factor is taken such as to enforce  $\dsp \int_{Y_{in}}^{Y_{out}} dy\int_0^{2h} dz  \phi_{Kl}^2 =1$. \\

The Fourier decomposition of $\phi_K$ can now be rewritten as $\dsp \phi_K(y,z)= \sum_{k,l\in \mathbb Z} a_{kl}\phi_{kl}(y,z)$.
Two modes corresponding to two different eigenvalues are orthogonal for the scalar product $\dsp (f|g) \equiv \int_{\mD}dydz\bar f g$. Hence,  
 $\phi_K$ is a solution of $\Delta^\star \psi = -\kappa^2 \phi_K$ \emph{iff} there exists $(k,l)$ such that  $\kappa_{kl}^2=\kappa^2$. 

\begin{figure}[htb]
 \centering
\includegraphics[width=0.5\textwidth]{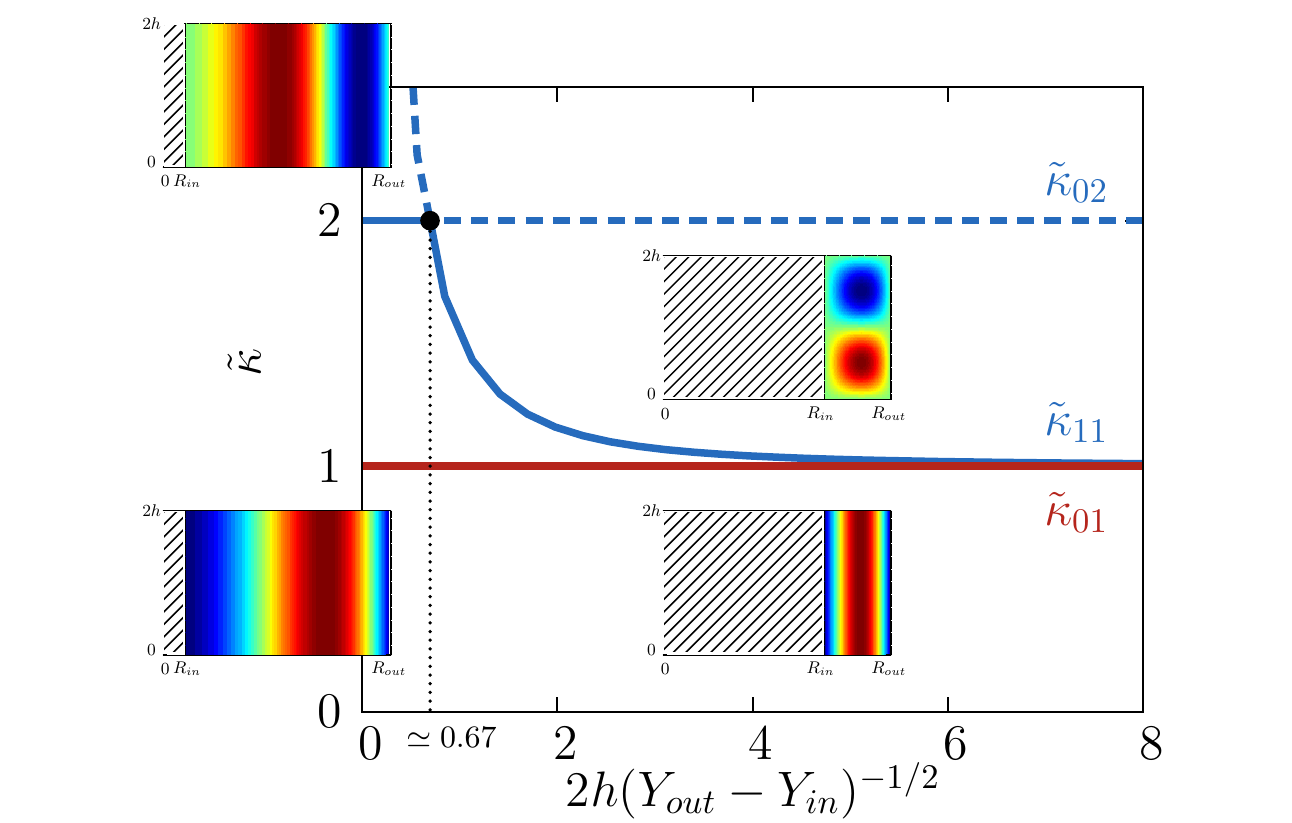} 
\caption{Numerical estimation of the first eigenvalues of $\Delta^\star$ as functions of the domain size. The eigenvalues $\kappa$ are adimensionnalised and $\tilde\kappa = \dfrac{\kappa}{\pi(Y_{out}-Y_{in})}$. The estimation was made with a fixed height $2h=1$ and fixed outer radius $R_{out}=\sqrt 2$. The inserted pictures represent maps of the corresponding eigenmodes.}
\label{fg:illus_vp}
 \end{figure}

\paragraph{}
As an illustration,  a numerical estimation for different domain shapes of the first eigenvalues of $\Delta^\star$ together with their corresponding eigenmode is provided on Figure \ref{fg:illus_vp}.

\subsection{Types of solutions for equation (\ref{eq:closure_reminder}).}

\paragraph{}
Let $\psi$ be a solution of equation (\ref{eq:closure_reminder}) and let us decompose $\psi$ as $\dsp \psi=\sum_{k,l}p_{kl}\phi_{kl}$.  Then necessarily, 
\begin{equation}
 \forall (k,l) \in {\mathbb Z}^2\,~~ p_{kl}\left(\kappa_{kl}^2+\dfrac{\beta^\star}{6}\right)=\dfrac{h^\star}{3}(1|\phi_{kl}).
 \label{eq:coeffs}
\end{equation}

Let us note that the only  modes with a non vanishing integral over the domain, -- namely such that $(1|\phi_{kl})\neq 0$ -- are the modes obtained for $k=0$ and $l$ odd.  To describe the solutions of equation (\ref{eq:coeffs}) we now need to consider the three following different cases. We hereby follow an existing terminology, as found for example in \cite{chavanis1996classification,naso2010statistical}.

 \paragraph{i) Continuum solutions.} If $\forall (k,l), \beta^\star \neq -6\kappa_{kl}^2$, then necessarily 
\begin{equation}
\forall (k,l) \,~ p_{kl}= \dfrac{h^\star(1|\phi_{kl})}{3\left(\kappa_{kl}^2+\dfrac{\beta^\star}{6}\right)}.
\end{equation}
 In this case, $\psi$ can be written as 
\begin{equation}
 \psi= \dfrac{h^\star}{3}\sum_{k,l}\dfrac{(1|\phi_{kl})}{\left(\kappa_{kl}^2+\dfrac{\beta^\star}{6}\right)} \phi_{kl} = \dfrac{h^\star}{3}\sum_{l \, odd}\dfrac{(1|\phi_{0l})}{\left(\kappa_{0l}^2+\dfrac{\beta^\star}{6}\right)} \phi_{0l} 
\end{equation}

For any odd value of $l$, this family of solution is continuous for values of $-\beta^\star/6$ between two eigenvalues $\kappa_{0l}^2$ and $\kappa_{0l+2}^2$ , and diverge for $-\beta^\star/6$ close to $\kappa_{0l}^2$. In particular, it is continuous for values of $-\beta^\star/6 =\kappa_{mn}^{2}$ such that $(1|\phi_{mn}) = 0$. 

 \paragraph{ii) Mixed solutions and eigenmodes.} Otherwise there exists $(k_0,l_0)$ such that $\beta^\star=-6 \kappa_{k_0l_0}^2$.  Then necessarily $\forall (k,l)\neq (k_0,l_0)\, ~p_{kl}=\dfrac{h^\star(1|\phi_{kl})}{3\left(\kappa_{kl}^2-\kappa_{0 l_0}^2\right)}$. 
 \subparagraph{ii.a) Mixed Solutions.} If $(1|\phi_{k_0l_0}) = 0$, -- \emph{e.g} if $k_0 \neq 0$ or $l_0$ is even --,  then $\psi$ can be written as $\dsp \psi = p_{k_0l_0}\phi_{k_0l_0} + \dfrac{h^\star}{3}\sum_{\substack{l \, odd}}\dfrac{(1|\phi_{0l})}{\left(\kappa_{0l}^2-\kappa_{k_0l_0}^2\right)} \phi_{0l}$. The coefficient $p_{k_0l_0}$ can take any value. $\psi$ can be seen as  a superposition of a solution from the continuum with the eigenmode $\phi_{k_0l_0}$, and we therefore call these solutions ``mixed solutions''.
 \subparagraph{ii.b) Odd eigenmodes.} Otherwise, $(1|\phi_{k_0l_0}) \neq 0$ -- \emph{e.g} $k_0 = 0$ and $l_0$ is odd. Equation (\ref{eq:coeffs}) considered for $(k,l) = (0,l_0)$ implies $h^\star=0$.  In this case $\psi$ is proportionnal to the odd eigenmode $\phi_{0l_0}$, namely $\psi= A \phi_{0l_0}$.

\section{Explicit derivation of the macrostate entropies}
We hereafter show how to derive the expressions (\ref{eq:poloidalentropy}) and (\ref{eq:KLfullmacroentropy}), which correspond to the critical macrostate poloidal entropy of the simplified problem, and the critical macrostate entropy of the full problem in the high energy regime.   

\subsection{Deriving the non-helical poloidal critical macrostate  entropies.}
Recall that the critical distributions $p_M^{\star,E}$ related to the non-helical poloidal problem are described by Equations (\ref{eq:reducedpfun}) and (\ref{eq:lagrangepol}). Recall that their macrostate entropy reads -- Equation (\ref{eq:polentropymacro}) -- :

\begin{equation}
\begin{split}
 \mS_M^\pol[p_M^{\star,E} ] &= -\dfrac{1}{\vD}\int_\mD \dx \int_{-M}^M \d\xi \,\pstarE \ln \pstarE \nonumber\\
 &= -\dfrac{1}{\vD}\int_\mD \dx \int_{-M}^M d\xi\,\pstarE \{\left(h\lM-\beta\lM\dfrac{\psi\x}{2}\right)\xi - \ln M -\ln Z_M^\star \x \} \nonumber\\
&= \ln M -\dfrac{1}{\vD} \left(h\lM X_\tot-\beta\lM E\right)  + \dfrac{1}{\vD}\int_\mD \ln Z_M^\star \x. 
\label{eq:spol_asymp}
\end{split}
\end{equation}

The last equality is obtained using $ \dsp \int_{-M}^M \d\, \xi\, \pstarE =1$ on one hand, and remembering that
\begin{equation*}
  \int_\mD \dx \int_{-M}^M \d\xi\, \pstarE = X_\tot  \text{~and~}   \int_\mD \dx \int_{-M}^M \d\xi\, \dfrac{\psi}{2}\xi\, \pstarE = E \text{~ on the other hand}. 
\end{equation*}
The asymptotic development of $ \ln Z_M^\star \x$ for large $M$ now yields 

\begin{align}
\ln Z_M^\star \x  \underset{M\to\infty}{=} & \ln \{2 + \int_{-1}^1d\xi \dfrac{\xi^2}{2M^2}\left(h^\star -\beta^\star\dfrac{\psi\x}{2} \right)^2 + o\left(\dfrac{1}{M^2}\right)\} \nonumber \\
 \underset{M\to\infty}{=} &\ln 2 + \dfrac{1}{6M^2}\left(h^\star -\beta^\star\dfrac{\psi\x}{2} \right)^2 + o\left(\dfrac{1}{M^2}\right).
\end{align}
Therefore,
\begin{equation}
\int_\mD\dx \ln Z_M^\star \x  \underset{M\to\infty}{=}\vD \ln 2 + \dfrac{1}{2M^2}\left( h^\star X_\tot-\beta^\star E\right) +  o\left(\dfrac{1}{M^2}\right)
\label{eq:zpol_asymp}.
\end{equation}

From  Equation (\ref{eq:spol_asymp}) and Equation (\ref{eq:zpol_asymp}), we finally obtain (\ref{eq:poloidalentropy}). 

\subsection{Deriving the (helical) critical macrostate  entropies in the high energy regime.}
\label{app:macroentropy}
For the full problem in the case of a non-vanishing poloidal energy, recall that the critical distributions $p_M^{\star}$ are given by Equation (\ref{eq:lagrangepol_unplugged}) and the scaling of the Lagrange multipliers by Equation  (\ref{eq:fullscaling}). In addition to the reduced Lagrange multipliers defined in (\ref{eq:fullscaling}), we also define  $\alpha_k^\star = \lim_{M\to\infty} M^2 (\alpha_k\lM-\alpha_k)$.\\

It is useful to express the Lagrange multipliers $h_k^\star$ and $\alpha_k$ in terms of the constraints. It is easily obtained from Equation (\ref{eq:fullmacroconstraints}) and Equation (\ref{eq:lagrangepol_unplugged}) that

\begin{equation}
 A_k = \vD \dfrac{\exp \alpha_k}{\sum_{{k^\prime}=1}^K \exp \alpha_{k^\prime}} \text{~~and~~} X_k = \dfrac{A_k h_k^\star}{3} - \dfrac{\beta A_k}{6 \vD} \int_\mD \dx \psi\x,
\label{eq:akxk}
\end{equation}

from which it follows that $\alpha_k = \ln \dfrac{A_k}{\vD}$ -- up to an unphysical  constant that can be absorbed in the partition function -- and $\dsp \dfrac{X_\tot}{\vD}-\dfrac{X_k}{A_k} = \dfrac{1}{3} \left( \overline{h_k^\star} -h_k^\star\right)$.\\

The critical points of the macrostate entropy then read
\begin{align}
 \mS_M[p_M^{\star} ] &= -\dfrac{1}{\vD}\int_\mD \dx \sum_{k=1}^K\int_{-M}^M \d\xi \,\pstark \ln \pstark \nonumber\\
  &= -\dfrac{1}{\vD}\int_\mD \dx \sum_{k=1}^K \int_{-M}^M d\xi\,\pstark \left\lbrace\alpha_k\lM -\beta\lM\dfrac{\sigma_k^2}{4y} \right.\\ & \hspace{3cm}\left. +\left(h_k\lM-\beta\lM\dfrac{\psi\x}{2}\right)\xi - \ln M -\ln Z_M^\star \x \right\rbrace \nonumber\\
&= \ln M -\dfrac{1}{\vD} \left(\sum_{k=1}^K \alpha_k\lM A_k + \sum_{k=1}^K h_k\lM X_k-\beta\lM E\right)  + \dfrac{1}{\vD}\int_\mD \ln Z_M^\star \x. 
\label{eq:dtildeasymp}
\end{align}

The last equality is obtained using $ \dsp \int_{-M}^M \d\, \xi\, \pstarE =1$ on one hand, and using Equation (\ref{eq:fullmacroconstraints}) to compute $A_k$, $X_k$, and $E$ on the other hand. The asymptotic development of $Z_M^\star \x$ for large $M$ then yields 
\begin{equation}
 Z_M^\star \x  \underset{M\to\infty}{=}  2 \sum_{k=1}^K e^{\alpha_k}\{1 +\dfrac{1}{M^2}\left[\alpha_k^\star -\beta^\star \dfrac{\sigma_k^2}{4y}+ \dfrac{1}{6}\left(h_k^\star -\beta^\star\dfrac{\psi\x}{2} \right)^2 \right] + o\left(\dfrac{1}{M^2}\right) \}.
\end{equation}
 Hence,
 \begin{equation}
\begin{split}
\int_\mD\dx \ln Z_M^\star \x  \underset{M\to\infty}{=} \vD  \ln 2 + \dfrac{1}{M^2} \left\lbrace \vD \overline{\alpha_k^\star} - \beta^\star E_\tor^\star  + \dfrac{1}{2}\sum_{k=1}^K h_k^\star X_k  \right. & \left. -\dfrac{\beta^\star}{2}E_\pol \right\rbrace \\ & +  o\left(\dfrac{1}{M^2}\right)
\end{split}
\label{eq:asymppart}.
\end{equation}

From  (\ref{eq:asymppart}) and  (\ref{eq:dtildeasymp}), we finally obtain    

\begin{equation}
 \mS_M [p_M^{\star}] \underset{M\to\infty}{=}  \ln 2M - \sum_{k=1}^K \dfrac{A_k}{\vD} \ln\dfrac{A_k}{\vD}+  \dfrac{1}{2\vD M^2}\left(\beta^\star E_\pol - \sum_{k=1}^K h_k^\star X_k \right) +  o\left(\dfrac{1}{M^2}\right),
\end{equation}
and equivalently the expression (\ref{eq:KLfullmacroentropy}).

\section{Maximizers of the macrostate entropy for the non-helical poloidal problem.}
\label{app:KLminimizers}
The constraints $E$ and $X_\tot$ being prescribed,  we want to determine  the values of $h^\star$ and $\beta^\star$ which minimize the poloidal macrostate entropy (\ref{eq:polentropymacro}).  We start from Equation (\ref{eq:poloidalentropy}). We want  to  determine  which among the critical distributions achieve the maximum of the macrostate entropy,  when $M$ is large. In the next paragraphs, we will rather work  with the reduced ``neg-entropy'' $D(\beta^\star,h^\star)$, whose minima are the maxima of the macrostate entropy : 

\begin{equation}
D(\beta^\star,h^\star) \underset{\text{def}}{=} \lim_{M \to\infty} \{-2M^2 \vD \mS_M^\pol [p_M^\star] +  \ln2M\}= \left( h^\star X_\tot-\beta^\star E\right).
\end{equation} 

\begin{figure}[htb]
 \centering
\includegraphics[width=0.49\textwidth,trim=0cm 0cm 2cm 0cm, clip]{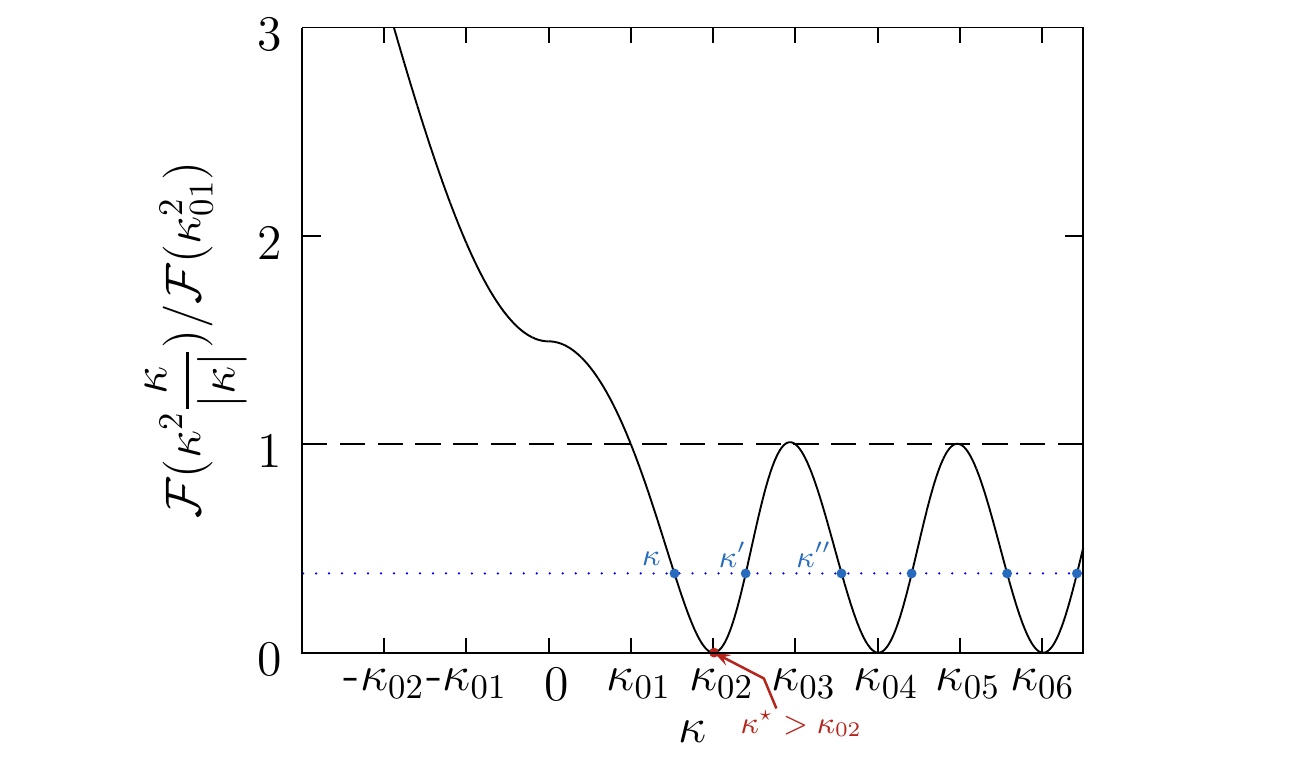} 
\includegraphics[width=0.49\textwidth,trim=0cm 0cm 2cm 0cm, clip]{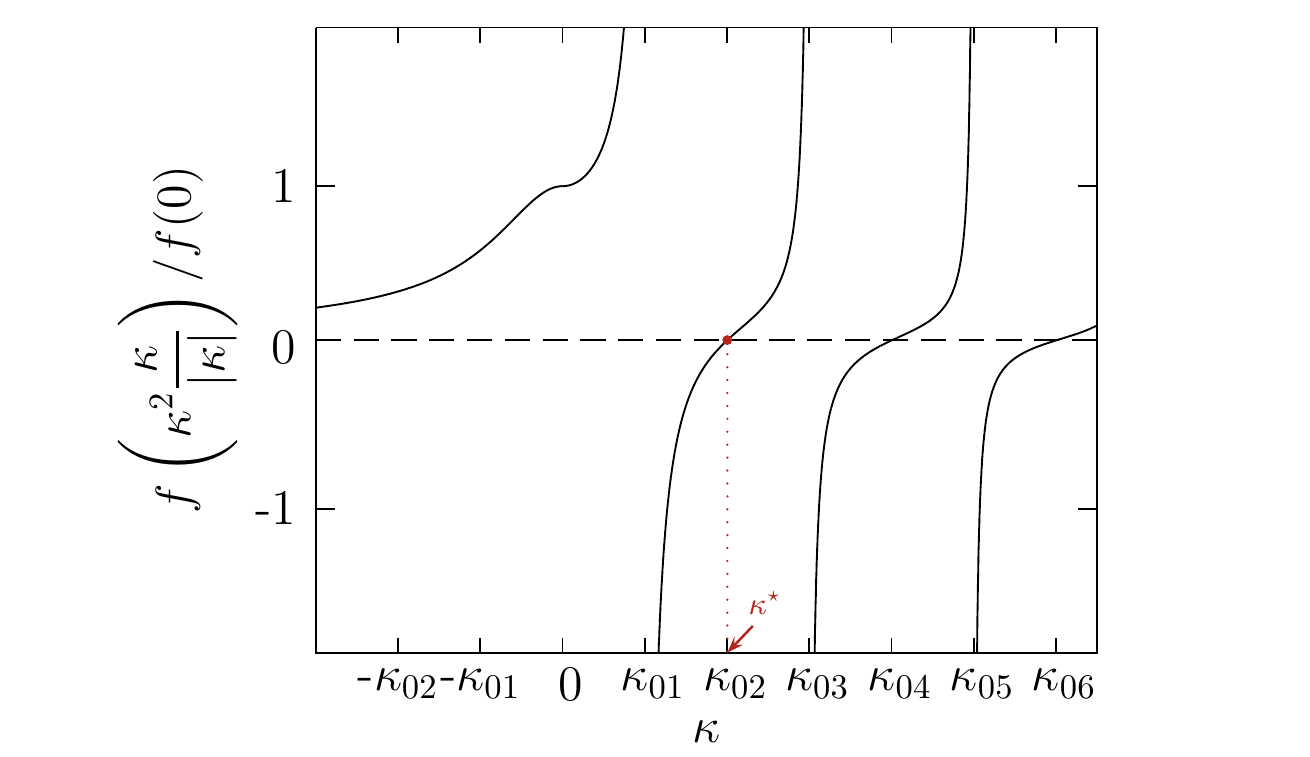} 
\caption{$\mathcal{F}$ and $f$ as  functions of $\kappa$. The minimum value $\kappa^\star$ for which both $\mathcal{F}$ and  $f$ are zero is greater than $\kappa_{02}$. }
\label{fig:ftilde}
 \end{figure}

It is convenient to define some  auxiliary functions :  
\begin{align}
&f(z) = \sum_{\substack{l \, odd}}\dfrac{(1|\phi_{0l})^2\kappa_{0l}^2}{\left(\kappa_{0l}^2-z\right) }, \text{~and~} \mF = \dfrac{f^2}{f^\prime}.
\end{align}
$f$ is defined on $\mathbb{R} -\{\kappa_{0(2l+1)}^2, l\in \mathbb{N}  \}$. $\mF$ is defined continuously over $\mathbb{R}$ by taking  $\mF(\kappa_{0l}) = (1|\phi_{0l})^2\kappa_{0l}^2= 16\pi /\vD$ for every odd value of $l$.
Those functions are sketched on Figure \ref{fig:ftilde}. 
We can  now relate $h^\star$ and $\beta^\star$ to $E$ and $X_\tot$ for each kind of solutions, in terms of $f$ and $F$ 

For a continuum solution,  
\begin{equation}
  X_\tot  = \dfrac{h^\star}{3} f\left(\dfrac{-\beta^\star}{6}\right), \text{~~}  2E=\dfrac{{h^\star}^2}{9} f^\prime\left(\dfrac{-\beta^\star}{6}\right), \text{~and~} X_\tot^2 = 2E \mF\left(\dfrac{-\beta^\star}{6}\right).
\label{eq:typei_hk}
\end{equation}

For a mixed solution, 
\begin{equation}
X_\tot = \dfrac{h^\star}{3} f\left(\kappa_{k_0l_0}^2\right), \text{~~}  2E=p_{k_0l_0}^2\kappa_{k_0l_0}^2 +\dfrac{{h^\star}^2}{9} f^\prime\left(\kappa_{k_0l_0}^2\right), \text{~and~} X_\tot^2 \le 2E \mF\left(\kappa_{k_0l_0}^2\right).\label{eq:type2a_hk}
\end{equation}

For an odd eigenmode, 
\begin{equation}
X_\tot^2  =2E_0 \kappa_{0l_0}^2(1|\phi_{0l_0})^2=2E_0\mF\left(\kappa_{0l_0}^2\right) \label{eq:type2b_hk}.
\end{equation}

It is clear from Figure \ref{fig:ftilde} and Equations (\ref{eq:typei_hk}), (\ref{eq:type2a_hk}) and (\ref{eq:type2b_hk}) that we need to make a distinction between the cases $\dfrac{X_\tot^2}{2E}>\mF(\kappa_{01}^2)$, $\dfrac{X_\tot^2}{2E}=\mF(\kappa_{01}^2)$, and $\dfrac{X_\tot^2}{2E}< \mF(\kappa_{01}^2)$. 

\subsection*{Case $\dfrac{X_\tot^2}{2E} > \mF(\kappa_{01}^2)$ }
\paragraph{}
In this case, the Lagrange multipliers $(h^\star,\beta^\star)$ are uniquely determined from the constraints. They describe  a solution from the continuum, which is therefore the maximal entropy solution. From a practical point of view, there is a one to one correspondance between the value of $\beta^\star$ and the value of $\dfrac{X_\tot^2}{2E}$ -- see Figure \ref{fig:ftilde}. We can therefore write without ambiguity $\beta^\star = -6 \mF^{-1}\left(\dfrac{X_\tot^2}{2E}\right)$. 
\spg
If $\dfrac{X_\tot^2}{2E} <\mF(0) $ , then $\beta^\star <0$ and we define $\kappa(\beta^\star) = \sqrt{-\beta^\star/6 }$. Otherwise, $\dfrac{X_\tot^2}{2E} \ge \mF(0)$, and $\beta^\star \ge 0$. We then define $\kappa (\beta^\star)= -\sqrt{\beta^\star/6}$. In both cases, $\kappa(\beta^\star)<\kappa_{01}$ and the other Lagrange multiplier is uniquely determined as $h^\star=\dfrac{3X_\tot}{f\left(-\beta^\star/6\right)}=\dfrac{3X_\tot}{f\left(\kappa^3/|\kappa|  \right)}$. 


\subsection*{Case $\dfrac{X_\tot^2}{2E} < \mF(\kappa_{01}^2)$ }
\paragraph{}
This case seems at first sight more intricated. First, there exist an infinite number of solutions from the continuum for which the constraints are satisfied. Indeed, for any odd value of $l$, there exist two values for the inverse temperature $\sqrt{-\beta^\star/6}$ in the interval $[\kappa_{0l};[\kappa_{0l+2}[ $ -- denoted by $\kappa$ and $\kappa^\prime$ on Figure \ref{fig:ftilde}. Second, there can exist an eigenvalue $\kappa_{k_0l_0}^2$ associated to an eigenmode $\phi_{k_0l_0}$ with $(1|\phi_{k_0l_0})=0$ such that  $\mF(\kappa_{k_0l_0}) > X_\tot^2/2E$. In this case, there also exists a mixed solution associated to the eigenvalue $\kappa_{k_0l_0}^2$ for which the constraints are satisfied.

The situation is however easily settled because the following result holds true. It is a non-trivial but fairly standard result \cite{chavanis1996classification}. 
\begin{result}
Between two solutions that satisfy the same constraints, the one associated with the lower value of $|\beta^\star|$ has the lower reduced neg-entropy -- and hence achieves the higher macrostate entropy.
\label{res:minimalentropy}
\end{result}

From the latter result, we deduce that if $\kappa_{\min}$ denotes the smallest eigenvalue whose associated eigenfunction has a vanishing mean on the domain, then
\begin{itemize}
  \item if $\mF(\kappa_{\min}^2) \le \dfrac{X_\tot^2}{2E}< \mF(\kappa_{01}^2)$, the selected solution is the solution from the continuum with inverse temperature $-6\beta^\star = \kappa^2 < \kappa_{\min}^2$ and $h^\star = 3 X_\tot /f(\kappa^2)$ uniquely determined from (\ref{eq:typei_hk}).
  \item if $\dfrac{X_\tot^2}{2E} \le \mF(\kappa_{\min}^2)$, the selected solution is the mixed solution, with inverse temperature satisfying $-6\beta^\star = \kappa_{\min}^2$ and $h^\star = 3 X_\tot /f(\kappa_{\min}^2)$ uniquely determined from (\ref{eq:type2a_hk}).
\end{itemize}

What remains to show is that (\ref{res:minimalentropy}) actually holds true. This is what the next two paragraphs are devoted to.

\paragraph{Maxima of the macrostate entropy achieved by the continuum solutions.}
Let us first focus on the continuum solutions. Those solutions are uniquely determined by the value of the inverse temperature $\beta^\star$. Indeed, from Equation (\ref{eq:typei_hk}), and given a value $\beta^\star$ such that $ \mF\left(-\beta^\star/6 \right)= \dfrac{X_\tot^2}{2E}$, then $h^\star$ is uniquely determined as $h^\star = 3 X_\tot /f(-\beta^\star/6)$. Defining $\kappa(\beta^\star)=\sqrt{-\beta^\star/6}$, we can write the reduced neg-entropy of such a continuum solution as 

\begin{equation}
 D^{\c}(\kappa(\beta^\star))= 6 \kappa(\beta^\star)^2 E+\dfrac{3X_\tot^2}{f\left(\kappa(\beta^\star)^2\right)}.
\label{eq:reducedconti}
\end{equation}

Let us now define  
\begin{equation}
\kappa= \min\{ \kappa^\prime \left| \mF \left({\kappa^\prime}^2 \right) \right. = \dfrac{X_\tot^2}{2E}\}.
\end{equation}

It is clear from Figure \ref{fig:ftilde} that $\kappa \in [\kappa_{01} ; \kappa^\star[$ where $\kappa^\star$ is the first zero of $\mF$. 

Then, $\kappa$ also achieves the minimal value of the reduced entropy (\ref{eq:reducedconti}), namely
\begin{equation}
  D^{c}\left(\kappa\right) = \min\{D^{c}\left( \kappa^\prime \right) \left| \mF \left({\kappa^\prime}^2 \right) \right. = \dfrac{X_\tot^2}{2E} \}.
\end{equation}

To see this, let $\kappa^\pprime >\kappa$ be such that $\mF({\kappa^\pprime}^2)= \mF({\kappa}^2)=\dfrac{X_\tot^2}{2E}$.
\begin{itemize}
\item If $f({\kappa^\pprime}^2)>0$, then 
  \begin{equation}
   D^{\c}(\kappa)-D^{\c}(\kappa^\pprime) = 6E\overbrace{\left(\kappa^2-{\kappa^\pprime} ^2\right)}^{<0} +3X_\tot^2 \overbrace{\left( \dfrac{1}{f(\kappa^2)}-\dfrac{1}{f({\kappa^\pprime}^2)}\right)}^{<0} <0. \label{eq:ineq1}
\end{equation}

\item Otherwise, let $\kappa^\prime = \sup\{\kappa | \kappa < \kappa^\pprime \text{~and~} \mF(\kappa^2)=\mF({\kappa^\pprime}^2) \}$. Then $f({\kappa^\prime}^2)>0$ (see Figure \ref{fig:ftilde}), and

\begin{equation}
D^{\c}(\kappa^\prime)-D^{\c}(\kappa^\pprime) < 6E{\left({\kappa^\prime}^2-{\kappa^\pprime} ^2\right)} +3X_\tot^2 \dfrac{{\kappa^\pprime}^2-{\kappa^\prime} ^2}{\mF({\kappa^\prime}^2)} \le 0.  \label{eq:ineq2}
\end{equation}
The first inequality of equation (\ref{eq:ineq2}) is obtained by using Tayor inequality at first order and by noticing that $(1/f)^\prime = -1/\mF$, while the second inequality stems froms the fact that $X_\tot^2=2E\mF({\kappa^\prime}^2)=2E\mF({\kappa^\pprime}^2)$. Therefore,
\begin{equation}
D^{\c}(\kappa)-D^{\c}(\kappa^\pprime)=D^{\c}(\kappa) - D^{\c}(\kappa^\prime)+ D^{\c}(\kappa^\prime)-D^{\c}(\kappa^\pprime) <0.
\end{equation}

\end{itemize}

\paragraph{Maxima of the macrostate entropy for continuum and mixed solutions.}
Let us now determine whether mixed solutions  can achieve a higher macrostate entropy than solutions from the continuum for the same prescribed contraints.  Consider for instance a mixed solution  associated to the eigenvalue $\kappa_{0}^2= \kappa_{k_0l_0}^2$. Equation (\ref{eq:type2a_hk}) tells that this solution exists provided $X_\tot^2 \le 2E \mF(\kappa_0^2)$. Let us suppose this is the case.
For this solution, the Lagrange multipliers are then uniquely determined as $\beta^\star=-6\kappa_0^2$, and $h^\star=\dfrac{3X_\tot}{f\left(\kappa_0^2\right)}$. The corresponding reduced neg-entropy reads  

\begin{equation}
D^{\m}(\kappa_0)=6\kappa_0^2 E + 3\dfrac{X_\tot^2}{f(\kappa_0^2)}. 
\end{equation}

We know from the previous paragraph, that the minimum of $D^{\c}(\kappa^\prime)$ is achieved for  some  $\kappa \in[\kappa_{01};\kappa_\star[$ which is uniquely determined.   We therefore need to compare $D^{\c}(\kappa)$ and $D^{\m}(\kappa_0)$.

\begin{itemize}
 \item If $\kappa_0 > \kappa_\star$, then inequalities similar to the inequalities (\ref{eq:ineq1}) and (\ref{eq:ineq2}) yield
$D^{\c}(\kappa) < D^{\m}(\kappa_0)$, so that the continuum solution f has a lower reduced neg-entropy and hence a higher macrostate entropy  than the mixed solution.
 \item Otherwise, we need to have  $\kappa_0 < \kappa < \kappa^\star$ in order for both solutions to exist. Then, 
\begin{equation}
 D^{\m}(\kappa_0)-D^{\c}(\kappa) \le 6E \left(\kappa_0^2 -\kappa^2\right)  + 3X_\tot^2\dfrac{\kappa^2-\kappa_0^2 }{\mF(\kappa^2) } < 0,
\label{eq:ineqi_ii}
\end{equation}
and the mixed solution has a lower reduced neg-entropy than any solution from the continuum that correspond to the same values of $E$ and $X_\tot$.
\end{itemize} 

Similar inequalities show that when two mixed solutions can coexist, it is the one associated with the lower value of $\kappa$ that also achieves the higher macrostate entropy.  

This concludes the proof of (\ref{res:minimalentropy}).
\subsection*{Case $\dfrac{X_\tot^2}{2E} = \mF(\kappa_{01}^2)$}
On this parabola, the only solutions that can exist are mixed solutions and pure odd mode solutions. 
For the odd eigenmodes, $h^\star=0$, the reduced entropy  simply reads $D^{\o}(\kappa_{0l})=6E \kappa_{0l}^2$. It is then clear, that the eigenmode with the lowest value of $D^{\o}$ is the gravest mode $\kappa_{01}$.

One can also notice that  $D^{\c}(\kappa_{01}+\epsilon) \underset{\epsilon \to 0 }{\to} D^{\o}(\kappa_{01})$. We can then extend by continuity Inequality (\ref{eq:ineqi_ii}), so that if there also exists a mixed solution  on the parabola $\dfrac{X_\tot^2}{ 2E} = \mF(\kappa_{01}^2)$,  it is the gravest odd mode that solves the extremization problem.

\subsection*{Conclusion}
We can now conclude the discussion. Recall that $\kappa_{\min}$ denotes the smallest eigenvalue with vanishing mean on the domain. Note that $\kappa_{\min}$ is lower than the first zero of $\mF$ (see Figure \ref{fig:ftilde}).
\begin{itemize}
 \item For $X_\tot^2 > 2E \mF(\kappa_{01}^2)$, the selected solution is a continuum solution, with $\kappa < \kappa_{01}$  uniquely determined by $E$ and $X_\tot$.
 \item For $X_\tot^2 = 2E \mF(\kappa_{01}^2)$, the selected solution is  the  gravest eigenmode $\kappa_{01}^2$.
 \item For $2E \mF(\kappa_{01}^2) >X_\tot^2 \ge 2E \mF(\kappa_{\min}^2)$, the selected solution is the one from the continuum associated to the value $ \kappa_{01}^2<\kappa^2 \le {\kappa^\star}^2$.
 \item For $2E \mF(\kappa_{\min}^2) \ge X_\tot^2 $ the selected solution is the mixed solution associated to the eigenvalue $\kappa_{\min}^2$.
\end{itemize}

\nocite{*}
\bibliographystyle{apalike}
\bibliography{biblio}
\end{document}